\documentclass[10pt,aps,prc,preprintnumbers,twocolumn,showpacs,showkeys,amsmath,floatfix,superscriptaddress]
{revtex4-1}
\usepackage{color,graphicx}
\usepackage{mathptmx}                
\usepackage{dcolumn}                 
\usepackage{bm}                      
\usepackage{ulem,soul}

\begin{document}

\preprint{INT-PUB-17-028}

\title{Density Functional approach for multi-strange hypernuclei: \\
competition between $\Lambda$ and $\Xi^{0,-}$ hyperons}

\author{J. Margueron}
\affiliation{Institute for Nuclear Theory, University of Washington, Seattle, Washington 98195, USA}
\affiliation{Institut de Physique Nucl\'eaire de Lyon, CNRS/IN2P3, Universit\'e de Lyon, Universit\'e Claude Bernard Lyon 1, F-69622 Villeurbanne Cedex, France}
\author{E. Khan}
\affiliation{Institut de Physique Nucl\'eaire, Universit\'e Paris-Sud, IN2P3-CNRS, 
Universit\'e Paris-Saclay, F-91406 Orsay Cedex, France}
\author{F. Gulminelli}
\affiliation{CNRS/ENSICAEN/LPC/Universit\'e de Caen Basse Normandy, 
UMR6534, F-14050 Caen Cedex, France}

\begin{abstract}
The question of the competition between $\Lambda$ and $\Xi^{0,-}$ in the ground-state of multi-strange hypernuclei is addressed
within a non-relativistic density functional approach, partially constrained by ab-initio calculations 
and experimental data.
The exploration of the nuclear chart for $10<Z<120$ as a function of the strangeness number is performed by adding hyperons to a nuclear core imposing either 
conserved total charge $Q$ or conserved proton number $Z$.
We find that almost all $\Lambda$ hypernuclei present an instability with respect to the strong interaction decay of $\Lambda$ towards $\Xi^{0,-}$ and that
most of the instabilities generates $\Xi^-$ (resp. $\Xi^0$) in the case of conserved total charge $Q$ (resp. proton number $Z$). 
The strangeness number at which the first $\Xi^{0,-}$ appear is generally lower for configurations explored in the case of conserved $Q$ compared to the case of conserved $Z$, and corresponds to the crossing between the $\Lambda$ and the neutron or proton chemical potentials.
About two to three hundred thousands pure $\Lambda$ hypernuclei may exist before the onset of $\Xi^{0,-}$.
The largest uncertainty comes from the unknown $\Lambda\Xi$ interaction, since the $N\Lambda$ and the $N\Xi$ ones can be constrained by a few experimental data.
The uncertainty on the $\Lambda\Xi$ interaction can still modify the previous estimation by 30-40\%, while the impact of the unknown $\Xi\Xi$ interaction is very weak.
\end{abstract}

\date{\today}


\maketitle

\section{Introduction}

{Since the discovery of the first hypernucleus in an emulsion exposed to cosmic rays~\cite{Danysz1953}, single and double-$\Lambda$ hypernuclei, as well as single-$\Xi$ ones, have been synthesised and some of their ground state properties have been measured~\cite{Fel2015,Gal2016}. }
It is further expected from theoretical calculations that multi-strange hyperons remain bound up to a large number of hyperons~\cite{Gal2016}, but precise predictions require reliable hyperon interactions.
The scarce amount of data however makes the hyperon interactions still rather unknown.
Depending on the hyperon interaction, hyperon might or might not appear in  dense matter -- hypernuclear matter -- which exists in the inner core of neutron stars~\cite{Schaffner2008}.
Finite hyper-nuclei and neutron stars are therefore the two systems which can provide constraints on the hyperon interactions.

New dedicated experimental programs such as Japan Proton Accelerator Research Complex (J-PARC) in Japan and the proton antiproton detector array at GSI Facility for Antiproton and Ion Research (FAIR) are or will be providing new data which participate to a better understanding of the properties of hypernuclei~\cite{Fel2015,Gal2016}.
The physics of hypernuclei opens a new direction in the exploration of the nuclear chart which is complementary to the direction towards more and more exotic nuclei. 
Hypernuclei are interesting finite nuclear systems since they allow to study the properties of bound strange hadrons and to test the behaviour of the baryon-baryon interaction.
The representation of the nuclear chart, traditionally expressed in terms of the number of protons $Z$ and neutrons $N$, 
acquires a new dimension associated to its strangeness number $S$.
For a given strangeness number $S$, several configurations corresponding to different hyperons can be considered. 
The charge neutral and lightest hyperon $\Lambda$ happens to be also the most bound, and single $\Lambda$ hypernuclei have  been synthesized through the nuclear chart, providing information such as global masses and single particle energies for most of them~\cite{has06}. These data are important to reduce the uncertainties of the $N\Lambda$ interaction, at least at very low density, as shown for instance in Refs.~\cite{Cugnon2000,Vidana2001}. 
Multi-strange hypernuclei are still one of the least-explored, open questions in hypernuclear physics, from both experimental and theoretical viewpoints~\cite{Ikeda1985}.
Data on double-$\Lambda$ hypernuclei are very scarce, mostly because the production rates are low. A few of them are nevertheless known, such as $^6_{\Lambda\Lambda}$He or $^{11}_{\Lambda\Lambda}$Be, allowing to extract the bond energy which is expected to be a measure of the $\Lambda\Lambda$ interaction, here also at very low density 
~\cite{Ahn13,Khan2015}.
The existence of an extra binding associated to a double-hyperon system implies that the $\Lambda\Lambda$ interaction is at least marginally attractive, opening the possibility of multi-strange systems with a higher number of hyperons~\cite{Kerman1974}. 
In particular, the production of multi-strange hypernuclei may be favoured during the cluster formation phase in relativistic heavy-ion collisions, since they usually lower the binding energy per particle~\cite{Kerman1974,Botvina2017}.

From the theoretical point of view, 
there have been many relativistic and non-relativistic density functional approaches which were developed and applied to the prediction of the structure of hypernuclei, see Refs.~\cite{Rayet1973,Bouyssy1982,Rufa1987,Glendenning1991,Cugnon2000,Vidana2001} for a few of them.
Multi-strange hypernuclei with more than two hyperons were first discussed in Ref.~\cite{Kerman1974}, and 
a large variety of phenomena have been predicted for such nuclei during the 80's and the early 90's~\cite{Rufa1987,Mares1989,Schaffner1992}. 
However, these studies 
assumed very attractive hyperon-hyperon interactions, inspired by the first analyses of double-Lambda $^{10}_{\Lambda \Lambda}$Be and $^{13}_{\Lambda \Lambda}$B data which suggested a large bond energy $\Delta B_{\Lambda\Lambda}\approx 5$ MeV~\cite{Franklin1995}. 
Therefore, it may be interesting to check these predictions with a density functional approach including the latest phenomenological constraints. Since our knowledge on the hyperon interaction remains quite poor, it may also be interesting to evaluate to which extent this lack of knowledge impacts the predictions of multi-strange hypernuclear properties.

In a previous work we have discussed hyperons and hypernuclear matter made of nucleons (N) and $\Lambda$ particles~\cite{Khan2015}.
It is however expected that multi-strange hypernuclei including also other hyperons, such as  $\Xi^{0,-}$, could be bound at large values of the strangeness number 
$S$~\cite{Dover1993,Schaffner1993,Balberg1994,Schaffner1994}.
For a given $S$, these complex configurations might even correspond to the ground state of the multi-baryon system, meaning that the hypernuclear chart at large $-S$~\cite{Khan2015} should take into account all the possible hyperons.
In this work, we want to investigate the properties of multi-strange hypernuclei with $\Lambda$ and $\Xi^{0,-}$ within non-relativistic density-functional theory which has proven to give a very good description of normal nuclei~\cite{cha98,ben03} and $\Lambda$-hypernuclei~\cite{Lanskoy1997, Lanskoy1998,Cugnon2000,Vidana2001,Zho2007,Minato2012,Schulze2013,Xian2016}, as well as $\Xi$-hypernuclei~\cite{Sun2016}.

Supposing an initial hypernucleus made of nucleons and $\Lambda$s,
 there are three kinds of possible strong interaction decays: i) reactions transforming $\Lambda$ into $\Sigma^{\pm,0}$,
\begin{eqnarray}
\Lambda + n \rightarrow \Sigma^0 + n, \hspace{0.1cm} \Lambda + p \rightarrow \Sigma^+ + n, \hspace{0.1cm}
\Lambda + n \rightarrow \Sigma^- + p,
\end{eqnarray}
with average free-reaction $Q_{\Sigma}^{free}\approx -80$~MeV; ii) reactions transforming two $\Lambda$ into $\Xi^{0,-}$,
\begin{eqnarray}
\Lambda + \Lambda \rightarrow \Xi^- + p, \hspace{0.5cm} \Lambda + \Lambda \rightarrow \Xi^0 + n,
\end{eqnarray}
with average free-reaction $Q_{\Xi}^{free}\approx -26$~MeV; and finally iii) reactions transforming three $\Lambda$ into $\Omega^-$,
\begin{eqnarray}
\Lambda + \Lambda + \Lambda \rightarrow \Omega^- + n + p, 
\end{eqnarray}
with average free-reaction $Q_{\Omega}^{free}\approx -180$~MeV.

The $Q^{free}$-values make the previously listed decays non-favorable, and the hypernuclei with only $\Lambda$ are usually preferred.
However, the $Q^{free}$-values takes into account only the mass of the particles, while in dense matter as well as in finite nuclei there is an additional quantum effect induced by Pauli blocking:
because of the Fermi energy, the total energy of hypernuclei with a large amount of $\Lambda$ may become larger than the total energy of a system where some $\Lambda$ are converted into other hyperons, leading to a positive $Q^{free}$-value in the medium. Moreover, the presence of other baryons in hypernuclei generates a potential field in non-relativistic approaches, or in-medium mass shift in relativistic approaches, which in turn shifts the $Q^{free}$-values.
The Coulomb interaction contributes also to the mean-field, and shifts it down for negatively charged hyperons, 
making them more favored.
Then, the minimum energy configuration for a fixed value of the quantum number set $(A,Q,S)$, may have a finite amount of $\Xi$, or $\Sigma$, or $\Omega$ particles.
The $Q^{free}$-values give however a reasonable hierarchy in the formation of new systems: it is expected that it will be easier to decay from $\Lambda$ to $\Xi$, than to $\Sigma$ and $\Omega$.

In this work, we therefore extend our previous analysis of the hypernuclear chart~\cite{Khan2015} considering the possible decay of $\Lambda$ into $\Xi^{0,-}$ (hereafter called the $\Xi$ instability).
The detailed study of the general properties of multi-strange hypernuclei is left to a future work. 
In the present work, we systematically look for the strangeness threshold associated to the appearance of $\Xi^{0,-}$ in the hypernuclear ground state. 
To calculate the $\Xi^{0,-}$ instability threshold, we consider a core-nucleus ($A_{core}$, $Z_{core}$) in between the drip-lines, and add  strangeness distributed 
over $\Lambda$ and $\Xi^{0,-}$ types of hyperons.
Fixing the three conserved charges of the strong interaction, namely the baryon number $A$, the total charge $Q$ and the strangeness number $S$, the ground state multi-strange hypernuclei is given by the one which minimizes the energy. 
This criterium corresponds to defining the stable configuration with respect to strong decays, 
and univocally defines the hypernuclear ground state.
To compare to some results in the literature, we also consider another convention: fixing $Z$ and adding strangeness on top of an core-nucleus ($A_{core}$, $Z_{core}$), as it was done for instance in Refs.~\cite{Dover1993,Schaffner1993,Balberg1994,Schaffner1994}.
It should be noted that a third strategy could be considered, which consists in adding strangeness at conserved total mass $A$, see Refs.~\cite{Rufa1987,Mares1989} for instance.
It should be stressed that  these strategies  are convenient
pictures to understand the effect of adding strangeness to ordinary nuclei, but none of them reflects exactly the present possibilities for the experimental production of multi-strange hypernuclei.
In particular in HIC, the reactions forming multi-strange hypernuclei can certainly produce extra excited states which do not correspond to the minimum energy at conserved $Q$. 

The outline of the present work is as follows.
In Sec.~\ref{sec:dft}, we propose a non-relativistic density-functional approach to treat multi-strange hypernuclei. 
We first briefly recall in Sec.~\ref{sec:func} the formalism already used in our previous work~\cite{Khan2015}, and propose in Sec.~\ref{sec:chi} a minimal extension to include the full baryonic octet with the inclusion of 10 additional coupling constants.
The multi-strange hypernuclear chart is studied in Sec.~\ref{sec:groundstates}.
 After a brief description of the numerical strategy (Sec.~\ref{sec:numerics}), the instability threshold corresponding to the onset of $\Xi^{0,-}$ hyperons in the ground state of hypernuclei is computed in Sec.~\ref{sec:decay} and the number of pure-$\Lambda$ hypernuclei is calculated in Sec.~\ref{sec:dripline}. 
We show that the use of realistic $N\Lambda$, $\Lambda\Lambda$ and $N\Xi$ interactions modifies the predictions with respect to previous results in the literature. 
The effect of the other largely unconstrained $YY$ couplings is also analysed and we show that the $\Lambda\Xi$ interaction channel is the most influential one.
Finally, conclusions and outlooks are presented in Sec.~\ref{sec:conclusions}.

\section{Density functional theory for multi-strange hypernuclei}
\label{sec:dft}

\begin{table*}[t]
\caption{Parameters of the f$_i$ functions, see Eqs.~(\ref{eq:fi1})-(\ref{eq:fi}), for the functionals DF-NSC89, DF-NSC97a, DF-NSC97f.}
\begin{ruledtabular}
\begin{tabular}{l|ccccccccccc}
Force & $\alpha_1^{N\Lambda}$ & $\alpha_2^{N\Lambda}$ & $\alpha_3^{N\Lambda}$ & $\alpha_4^{N\Lambda}$ & $\alpha_5^{N\Lambda}$ & $\alpha_6^{N\Lambda}$ & $\alpha_1^{\Lambda\Lambda}$ & $\alpha_2^{\Lambda\Lambda}$ & $\alpha_3^{\Lambda\Lambda}$ \\
& MeV~fm$^3$ & MeV~fm$^6$ & MeV~fm$^9$ 
& MeV~fm$^{5/3}$ & MeV~fm$^{14/3}$ & MeV~fm$^{23/3}$ 
& MeV~fm$^3$ & MeV~fm$^6$ & MeV~fm$^9$ \\
\hline
DF-NSC89~\cite{Cugnon2000,Vidana2001} & 327 & 1159 & 1163 & 335 & 1102 & 1660 & 0 & 0 & 0 \\
DF-NSC97a~\cite{Vidana2001} & 423 & 1899 & 3795 & 577 & 4017 & 11061 & 38 & 186 & 22 \\
DF-NSC97f~\cite{Vidana2001} & 384 & 1473 & 1933 & 635 & 1829 & 4100 & 50 & 545 & 981 \\
\end{tabular}
\end{ruledtabular}
\label{table:SZ1}
\end{table*}

In the present work, we consider the most general non-relativistic system composed of interacting nucleons $N$ (neutrons and protons) and hyperons, hereafter noted $Y$ for
 $\Lambda$, and $\Xi^{0,-}$.
Notice that the extension to the other hyperons, $\Sigma^{0,\pm}$, and eventually $\Omega^-$, is straight-forward, but will not be considered in this paper.
The total Hamiltonian reads,

\begin{eqnarray}
\hat{H}
&=&\sum_{i=N,Y} \hat t_i + \sum_{i,j=N,Y} \hat{v}_{ij}^{NY} + \frac{1}{2} \sum_{i=N} \hat{v}_{ii}^{NN}
+ \frac{1}{2} \sum_{i=Y} \hat{v}_{ii}^{YY}
\label{ham2} .
\end{eqnarray}

In the following, we will consider the density functional theory
which allows relating in a direct way the microscopic Brueckner-Hartree-Fock (BHF) theory for uniform matter based on the Nijmegen interactions, to the properties of hypernuclei.

\subsection{Energy-density functional for N and $\Lambda$ hypernuclear matter}
\label{sec:func}

In a previous study of hypernuclei and nuclear matter~\cite{Khan2015} we used a density functional which  was determined
directly from the BHF theory including nucleons and single $\Lambda$-hyperons~\cite{Cugnon2000,Vidana2001}.  
Here we recall the main equations and refer the reader to Ref.~\cite{Khan2015} for more details.

The total energy density $\epsilon(\rho_N,\rho_\Lambda)$ is related to the energy per particle of infinite nuclear matter calculated within
the BHF framework, $e_{BHF}$, as $\epsilon(\rho_N,\rho_\Lambda)=(\rho_N+\rho_\Lambda) e_{BHF}(\rho_N,\rho_\Lambda)$
and is decomposed in different terms,
\begin{eqnarray}
\epsilon(\rho_N,\rho_\Lambda) &=& \frac{\hbar^2}{2m_N}\tau_N+ \frac{\hbar^2}{2m_\Lambda}\tau_\Lambda 
+\epsilon_{NN}(\rho_N)  \nonumber \\
&&+\epsilon_{N\Lambda}(\rho_N,\rho_\Lambda)
+F_{\Lambda}\epsilon_{\Lambda\Lambda}(\rho_\Lambda) ,
\label{functional}
\end{eqnarray}
where $\tau_N$ and $\tau_\Lambda$ are  the kinetic energy densities,  
and the term $\epsilon_{N\Lambda}$ is parameterized in terms of the nucleon and hyperon densities
as~\cite{Cugnon2000,Vidana2001},
\begin{eqnarray}
\epsilon_{N\Lambda}(\rho_N,\rho_\Lambda) &=& -f_1^{N\Lambda}(\rho_N) \rho_N\rho_\Lambda + f_2^{N\Lambda}(\rho_N)\rho_N\rho_\Lambda^{5/3},
\label{eq:enl}
\end{eqnarray}
Here the first term physically corresponds to the attractive $N\Lambda$ interaction,
corrected by the presence of the medium given by the function
$f_1$, and the second term is induced by the repulsive momentum dependent term
of the $\Lambda$ potential (considering the low-momentum quadratic
approximation), also corrected by the medium through the function
$f_2$. 
Some repulsion is indeed necessary at high density for the $\Lambda$ in nuclear matter, as fits to single $\Lambda$ hypernuclear data have revealed~\cite{Millener1988}.
In the presence of the attractive $\Lambda\Lambda$ interaction,
the term $\epsilon_{\Lambda\Lambda}$ is solely determined by the
hyperon density as~\cite{Vidana2001},
\begin{equation}
\epsilon_{\Lambda\Lambda}(\rho_\Lambda)=-f^{\Lambda\Lambda}(\rho_\Lambda)\rho_\Lambda^2 .
\label{e_LL}
\end{equation}
To avoid self-interaction, the factor $F_{\Lambda}$ in the functional Eq.~(\ref{functional}) is 0 if there is only one $\Lambda$ and 1 for more.

The functions $f$ are given by the polynomial forms,
\begin{eqnarray}
f_1^{N\Lambda}(\rho_N) &=& \alpha_1^{N\Lambda}-\alpha_2^{N\Lambda}\rho_N+\alpha_3^{N\Lambda}\rho_N^2 , \label{eq:fi1}\\
f_2^{N\Lambda}(\rho_N) &=& \alpha_4^{N\Lambda}-\alpha_5^{N\Lambda}\rho_N+\alpha_6^{N\Lambda}\rho_N^2 , \\
f^{\Lambda\Lambda}(\rho_\Lambda) &=&\alpha_1^{\Lambda\Lambda}-\alpha_2^{\Lambda\Lambda}\rho_\Lambda+\alpha_3^{\Lambda\Lambda}\rho_\Lambda^2.
\label{eq:fi}
\end{eqnarray}

The values for the parameters $\alpha_1^{N\Lambda}$-$\alpha_6^{N\Lambda}$ were determined in Refs.~\cite{Cugnon2000,Vidana2001} from a fit of the BHF infinite nuclear matter calculations performed with different $N\Lambda$ potentials  which equally well fit the available $N\Lambda$ phase shifts. In this work  we will use the models DF-NSC89, DF-NSC97a and DF-NSC97f, which parameters are given in Tabs.~\ref{table:SZ1} and \ref{table:SZ2}.

It should be noted that no direct experimental information is available on $\Lambda\Lambda$ scattering, meaning that these phenomenological bare interactions are
rather unconstrained in the $\Lambda\Lambda$ channel. 
For this reason, NSC89 does not contain any $\Lambda\Lambda$ interaction.
The NSC97a-f models assume for this channel a simple SU(3) extension of the original Nijmegen potential models to multiple strangeness $S=-2$~\cite{Maessen1989,Stoks1999}.
For these models, the energy density associated to the $\Lambda\Lambda$ interaction is expressed as
\begin{equation}
\epsilon_{\Lambda\Lambda}=-\left(\alpha_1^{\Lambda\Lambda}-\alpha_2^{\Lambda\Lambda}\rho_\Lambda+\alpha_3^{\Lambda\Lambda}\rho_\Lambda^2\right)\rho_\Lambda^2 .
\label{eq:tildeeLLunif}
\end{equation}
It turns out that these models do not lead to a satisfactory description of the bond energy of double-$\Lambda$ hypernuclei~\cite{Vidana2001}, which is the only empirical information that we have on $\Lambda\Lambda$ couplings~\cite{Franklin1995,Aoki09,Ahn13}. 
For this reason, in our previous work in Ref.~\cite{Khan2015}, we have empirically modified the $\alpha_1^{\Lambda\Lambda}$-$\alpha_3^{\Lambda\Lambda}$ parameters such as to reproduce the measured binding energy of $^6_{\Lambda\Lambda}$He.
It should be noted that the parameters $\alpha_2^{\Lambda\Lambda}$-$\alpha_3^{\Lambda\Lambda}$ control the high density behavior of the $\Lambda\Lambda$
interaction. 
In that same work, we found that the global properties of $\Lambda$-hypernuclei were not impacted by the high density behavior of the $\Lambda\Lambda$ interaction~\cite{Khan2015}.
Therefore the parameters $\alpha_2^{\Lambda\Lambda}$-$\alpha_3^{\Lambda\Lambda}$ have no impact in double-$\Lambda$ hypernuclei, since the $\Lambda$ density in these systems remains rather small.
In the present work, by including additional hyperons, the $\Lambda$-density in multi-Y hypernuclei is expected to be even further reduced compared to the case of pure $\Lambda$-hypernuclei.
We therefore simplify the $\Lambda\Lambda$ interaction as expressed in Eq.~(\ref{e_LL}), to its first term as,
\begin{equation}
{\epsilon}_{\Lambda\Lambda}=-
{\alpha}_1^{\Lambda\Lambda}\rho_\Lambda^2 .
\label{eq:simple}
\end{equation}
In the following, we will refer to the modification of the $\Lambda\Lambda$ interaction Eq.~(\ref{eq:simple})
as the EmpC prescription.

The parameter ${\alpha}_1^{\Lambda\Lambda}$ can still be approximately related to the average bond energy expected from a local density approximation
$\Delta B_{\Lambda\Lambda}$ and the average density of $\Lambda$ inside the nucleus, $x_\Lambda=\rho_\Lambda/\rho_0$ (see Ref.~\cite{Khan2015} for details), as
\begin{equation}
{\alpha}_1^{\Lambda\Lambda}=\frac 1 2 \frac{\Delta B_{\Lambda\Lambda}}
{\rho_0 x_\Lambda} .
\label{eq:bondLDA}
\end{equation}
A recent publication questions the validity of the local density approximation~\cite{Fortin2017}, and points out the fact that the $\Lambda$ potential obtained
imposing Eq.~(\ref{eq:bondLDA}) with a constant value $x_\Lambda=1/5$ depends on the chosen functional, and so does the corresponding bond energy obtained by a direct HF hypernuclear calculation.  
If however the value of $x_\Lambda$ is consistently obtained for each interaction model by a self-consistent HF calculation, we have shown in Ref.~\cite{Khan2015} that this simple prescription leads to very precise results.
This point is demonstrated for the EmpC prescription in Table~\ref{table:presC}, which  shows 
the final values for ${\alpha}_1^{\Lambda\Lambda}$ for SLy4 and the $\Lambda$N potentials, DF-NSC89, DF-NSC97a and NSC97f,
together with the HF results for the bond energy and the ratio of the average $\Lambda$-density to the saturation density in He obtained from our Hartree-Fock calculations. 
The resulting bond energy is very close to the value $\Delta B_{\Lambda\Lambda}=1$ MeV imposed by Eq.~(\ref{eq:bondLDA}), provided the consistent value of $x_\Lambda$ obtained in the $^6_{\Lambda\Lambda}He$ ground state, and given in Table~\ref{table:presC}, is used.

In the nucleon sector, we use the SLy4 parametrization of the phenomenological Skyrme functional including non-local and spin-orbit terms, since it  is the NN interaction which has been used to calibrate the $\Lambda\Lambda$ interaction~\cite{Vidana2001}, and it can correctly reproduce the properties of stable and exotic nuclei~\cite{cha98}. 

{A three body YNN repulsive interaction has recently been proposed in relation with the hyperonization puzzle~\cite{Lonardoni2015}: usual NN and NY interactions fitted on phase shifts cannot predict neutron star masses above about 1.6~$M_\odot$, while such objects were recently observed \cite{DeMorest,Antoniadis}.
This YNN repulsive interaction was originally introduced to improve the agreement between the experimental $\Lambda$-separation energies and the one predicted from an Argonne like two-body potential~\cite{Lonardoni2013}. 
It is interesting to remark that the functional we use does not include a bare NNY interaction, and still is able to well reproduce the experimental $\Lambda$-separation energies \cite{Cugnon2000,Vidana2001}. This seemingly contradictory result can be qualitatively explained by the fact that 
the Nijmegen functional   (\ref{eq:enl}) has indeed a two-body induced NNY term, the $\alpha_2^{N\Lambda}$ term.
It would however be interesting to have a functional form adjusted to the ab-initio calculations reported in Refs.~\cite{Lonardoni2013,Lonardoni2015} for future systematic applications to multi-hypernuclei such as in this work.

\begin{table}[b]
\renewcommand{\arraystretch}{1.3}
\caption{ Prescription EmpC. We present the values of the parameters ${\alpha}_1^{\Lambda\Lambda}$, 
the resulting bond energy in He $\Delta B_{\Lambda\Lambda}(A=6)$~ in MeV,
 and the ratio of the $\Lambda$-density in He to the
saturation density ($\rho_0$).}
\begin{ruledtabular}
\begin{tabular}{ccccccc}
Pot. $\Lambda$N & DF-NSC89 & DF-NSC97a & DF-NSC97f  \\
Pot. $\Lambda\Lambda$ & EmpC & EmpC & EmpC \\
\hline 
${\alpha}_1^{\Lambda\Lambda}$ (MeV~fm$^3$)& 22.81 & 21.12 & 33.25  \\
$\Delta B_{\Lambda\Lambda}(6)^{HF}$ (MeV) & 1.00 & 0.99 & 1.01 \\
$\rho_\Lambda(6)/\rho_0$ & 0.137 & 0.148 & 0.094 \\
\end{tabular}
\end{ruledtabular}
\label{table:presC}
\end{table}

\subsection{Generalization of the energy functional for N and multi-Y hypernuclear matter}
\label{sec:chi}

In this section, we propose a general and simple density-functional considering the full hyperon octet.
The functional form~(\ref{functional}) is generalized in order to include $Y=\Lambda$ and $\Xi^{0,-}$ hyperons, as
\begin{eqnarray}
\epsilon &=& \frac{\hbar^2}{2m_N}\tau_N +\epsilon_{NN}(\rho_N)  \nonumber \\
&+& \sum_{Y}\frac{\hbar^2}{2m_Y}\tau_Y
+\sum_{Y}\epsilon_{NY}(\rho_N,\rho_Y)
\nonumber \\
&+& \sum_{Y_1,Y_2}\epsilon_{Y_1Y_2}(\rho_{Y_1},\rho_{Y_2}) 
+\sum_{Y}F_{Y}\epsilon_{YY}(\rho_Y) .
\label{functionaltot}
\end{eqnarray}
Using the isospin invariance of the strong interaction and neglecting spin dependence for simplicity, we suppose that the general-functional~(\ref{functionaltot}) depends only on the following densities
$\rho_N$, $\rho_\Lambda$, and $\rho_\Xi=\rho_{\Xi^0}+\rho_{\Xi^-}$.
It should be noted that while the spin-orbit interaction between N and $\Lambda$ is known to be small~\cite{Finelli09}, it is not certain that it is also the case
for the interaction channels involving other hyperons.
Due to the lack of data to set the spin-orbit interaction strength, we neglect the spin-orbit interactions in all $NY$ and $YY$ channels in the present work.
In Eq.~(\ref{functionaltot}), the parameters $F_Y$ are  introduced in finite systems to avoid self-interactions: $F_Y=1$ if the associated number of $Y$, $N_Y\ge 2$, $F_Y=0$ otherwise.

\begin{figure}[tb]
\begin{center}
\includegraphics[width=0.5\textwidth]{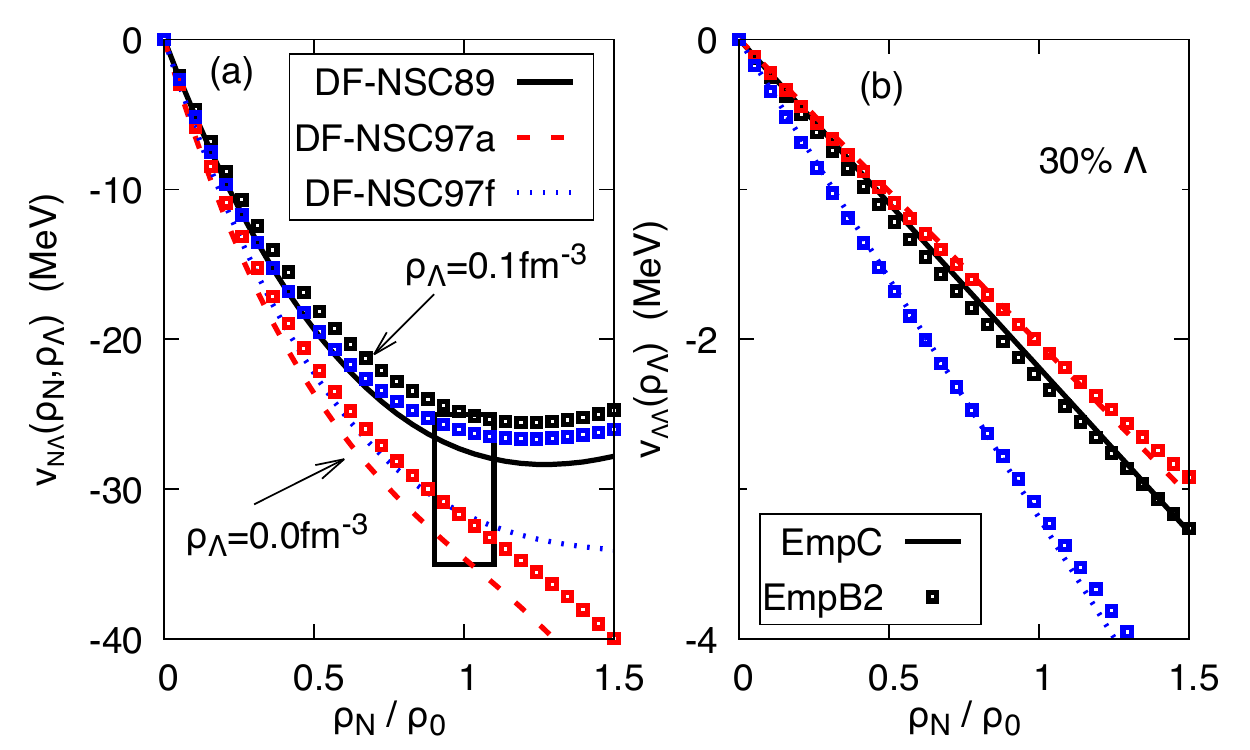}
\caption{(Color online) Potentials $v_{N\Lambda}(\rho_N,\rho_\Lambda)$ (a) and $v_{\Lambda\Lambda}(\rho_\Lambda)$ (b) as a function of the nucleon density $\rho_N$ in units of the saturation density 
$\rho_0$, for the functionals DF-NSC89, DF-NSC97a, and DF-NSC97f. In panel (a) two different $\Lambda$ densities are considered: 
$\rho_\Lambda=0.0$~fm$^{-3}$ (solid lines) and $\rho_\Lambda=0.1$~fm$^{-3}$ (empty squares). In panel (b) the simplified prescription EmpC (solid lines) is compared to the prescription EmpB2 from Ref.~\cite{Khan2015} (empty squares) in the case when 30\% of the baryons are taken as $\Lambda$.}
\label{fig:vl}
\end{center}
\end{figure}

The mean-field potentials are deduced from the functional~(\ref{functionaltot}) by using functional derivative~\cite{Book:RingSchuck}.
Fig.~\ref{fig:vl} displays the $N\Lambda$ and the $\Lambda\Lambda$ mean fields defined as $v_{XY}^{unif}=\partial\epsilon_{XY}/\partial\rho_Y$
for the functionals DF-NSC89, DF-NSC97a and DF-NSC97f.
On the left panel the $N\Lambda$ potential is shown for two cases: in the absence of $\Lambda$ and for $\rho_\Lambda=0.1$~fm$^{-3}$.
On the right panel, 30\% of the baryons are taken as $\Lambda$.
The $N\Lambda$ mean field is consistent with the empirical expectation (box on Fig.~\ref{fig:vl}), and a finite amount of $\Lambda$ decreases the depth of the mean field, as expected~\cite{Millener1988}.
There is a qualitative difference between the functional DF-NSC97a and the two others for a small amount of $\Lambda$: the functional DF-NSC97a is much more
attractive than the two others, which are rather equivalent.
We will see in the following that hypernuclei predicted by DF-NSC97a are consequently more bound.
On the right panel, the prescriptions EmpB2 from Ref.~\cite{Khan2015} and the present simplified functional quadratic in density EmpC are compared for the 
$\Lambda\Lambda$ channel.
It is shown that up to a large amount of $\Lambda$ (30\%) and at saturation density, there is almost no difference between the two prescriptions EmpB2 and EmpC for the $\Lambda\Lambda$ interaction.
This implies that the parameters $\alpha_2^{\Lambda\Lambda}$-$\alpha_3^{\Lambda\Lambda}$ can indeed be ignored for hypernuclei as discussed above, and that
the density dependence of the $\Lambda\Lambda$ interaction cannot be constrained by hypernuclear physics, as concluded in our previous study~\cite{Khan2015}.

To study this feature in more details, Fig.~\ref{fig:lhn} displays the evolution of the binding energies for a few illustrative nuclei as a function of the strangeness number $-S$: $^{40-S\Lambda}$Ca, $^{56-S\Lambda}$Ni, $^{120-S\Lambda}$Sn, $^{208-S\Lambda}$Pb.
The calculations are stopped at the $\Xi$-instability.
The similarity in the predictions for the $\Lambda\Lambda$ interactions given by EmpB2 and its simplified version EmpC is demonstrated in Fig.~\ref{fig:lhn}.
This figure clearly shows that the high density contribution of the $\Lambda\Lambda$ interaction, namely the parameters 
$\alpha_2^{\Lambda\Lambda}$-${\alpha}_3^{\Lambda\Lambda}$ in Eq.~(\ref{eq:fi}) which are neglected in the present parametrization EmpC, has no contribution to the mean field in multi-$\Lambda$-hypernuclei.
This is consistent with our previous conclusions in Ref.~\cite{Khan2015} and confirms that the hyperon density in hypernuclei remains too low to provide information on the hyperon matter above saturation density.

\begin{figure}[tb]
\begin{center}
\includegraphics[width=0.5\textwidth]{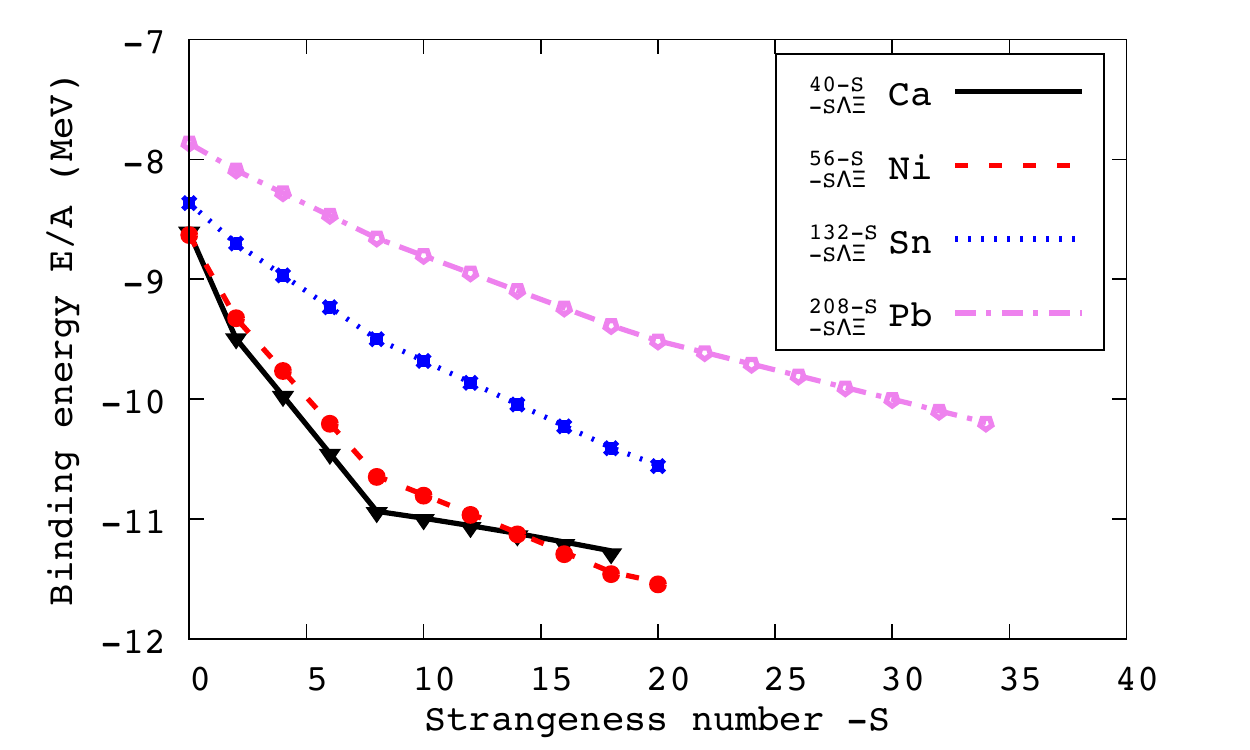}
\caption{(Color online) Binding energy $E/A$ at conserved $Q$ for different multi-$\Lambda$-hypernuclei function of the strangeness number $-S$:
$^{40-S\Lambda}$Ca, $^{56-S\Lambda}$Ni, $^{120-S\Lambda}$Sn, $^{208-S\Lambda}$Pb.
We compare predictions from two functionals for the $\Lambda\Lambda$ interaction, DF-NSC89+EmpC (lines) and DF-NSC89+EmpB2 (symbols),
using SLy4 for the $N$ interaction and VY0 for the $\Xi$ interaction. The calculations are stopped at the $\Xi$-instability threshold.}
\label{fig:lhn}
\end{center}
\end{figure}

As expected from previous works (see Ref.~\cite{Gal2016} and references therein) the binding energy increases in absolute value when hyperons
are added to normal nuclei (see Fig.~\ref{fig:lhn}) and for this reason it was suggested that  multi-$\Lambda$ hypernuclei may be formed in heavy-ion collisions~\cite{Schaffner1992}. 
As we have stressed in the introduction, this can be understood from the fact that adding hyperons opens new degree of freedom for which the Fermi energy is small. 
As the number of $\Lambda$ increases this effect is less and less important, and finally, the binding energy saturates with the number of $\Lambda$, for a large number of $\Lambda$.
For an even larger number of $\Lambda$, the total binding energy increases, and thus the chemical potential $\mu_\Lambda$ becomes positive indicating that the 
$\Lambda$ drip-line is met (and the calculation is stopped at that point).

For a large number of $\Lambda$ one might expect that it could again be energetically favoured to add new types of hyperons, such as the $\Xi^{0,-}$.
It should be noted however that when we are comparing different possible ground states configurations conserving $A$, $S$ and $Q$ (or $Z$), 
the final result might not be easy to anticipate because it depends on different competing effects.
First, the higher  masses of  $\Xi^{0,-}$, $\Sigma^{0,\pm}$ and $\Omega^-$ bring a penalty for the onset of an addition of these hyperons (see the discussion of the $Q^{free}$-values in the introduction of this work).
In addition, the interaction of these new hyperons with the nucleons, which is the dominant contribution, might be attractive or repulsive, impacting their mean-field potential.
The attractive Coulomb interaction for negatively charged hyperons could help the binding, as we will show in Sec.~\ref{sec:groundstates}. 
Finally, the higher mass also induces a reduction of the kinetic energy of these particles,  which could therefore slightly counterbalance the effect of a weaker interaction.
All these phenomena are naturally included in our framework and in the following, we will study all their combined effects in more details.

We now discuss the $\Xi$ channel.
This channel is much less known than the $\Lambda$ one.
The $\Xi$ density is expected to remain quite low, even lower than the $\Lambda$-density in the case of pure $\Lambda$-hypernuclei~\cite{Khan2015}, since they shall be less numerous than the $\Lambda$, which is expected to be the more bound hyperon.
In addition, the effective masses of the $\Xi$ is assumed equal to their bare masses.
This is justified from recent Bruckener-Hartree-Fock calculations~\cite{Schulze2013,Rijken2016} for the $\Xi$ and it was also assumed
in recent density-functional approaches~\cite{Sun2016}.
Indeed, if we are only interested in ground-state properties, the effective masses can be incorporated in a deeper mean-field. 
Since we do not know much concerning the mean field or the effective masses of these hyperons, it is simpler to assume the effective mass equal to the bare mass, and eventually alter the depth of the mean-field. 

Following the previous arguments,  the  $NY$ ($Y=\Xi^{0,-}$)  terms of the potential energy density functional are given  a quadratic density dependence, as obtained for a simple two-body effective interaction: 
\begin{equation}
\epsilon_{NY}(\rho_N,\rho_Y) = -\alpha^{NY} \rho_N\rho_Y ,
\end{equation}
and the same is assumed for the  $YY^\prime$ terms ($Y^\prime=\Lambda$ and $\Xi^{0,-}$):  
\begin{equation}
\epsilon_{YY^\prime}(\rho_Y,\rho_{Y^\prime})=-\alpha^{YY^\prime} \rho_Y\rho_{Y^\prime} ,
\end{equation}
leading to the definition of three additional constants. 

\begin{table}[t]
\renewcommand{\arraystretch}{1.3}
\begin{ruledtabular}
\begin{tabular}{ccccccccccc}
$\alpha_1^{N\Xi}$  & $^{12}_{\Xi s}$Be & $^{15}_{\Xi s}$C & $^{15}_{\Xi p}$C \\
\hline
\multicolumn{4}{c}{Interpretation 1 of the "Kiso" event} \\
105 & 2.64 & 3.92 & 0.12 \\
109 & 3.05 & 4.36 & 0.29 \\
110 & 3.16 & 4.47 & 0.34 \\
Exp. & $\approx$5$^\dagger$ & 4.38$\pm$0.25 & 1.11$\pm$0.25 \\
Ref.  & \cite{Khaustov2000,Hiyama2008} & \cite{Gogami2016} & \cite{Nak2015} \\
\hline
\multicolumn{4}{c}{Interpretation 2 of the "Kiso" event} \\
120   & 4.23 & 5.60 & 0.84 \\
125   & 4.79 & 6.18 & 1.13 \\
130   & 5.36 & 6.78 & 1.81 \\
Exp. & $\approx$5$^\dagger$ & 7.2-9.4$^\dagger$ & 1.11$\pm$0.25 \\
Ref.  & \cite{Khaustov2000,Hiyama2008} & \cite{Sun2016} & \cite{Nak2015} \\
\end{tabular}
$^\dagger$ Theoretical expectation.
\end{ruledtabular}
\caption{The $\Xi^-$ removal energies $B_{\Xi^-}$ (in MeV, see Eq. \ref{eq:kiso}) of $^{12}_{\Xi s}$Be in its ground-state and
of $^{15}_{\Xi s}$C in its ground and first excited states. Interpretations 1 and 2 stands for the two 
possible interpretations of the "Kiso" event~\cite{Nak2015} (see text).}
\label{table:removal}
\end{table}

The parameter $\alpha^{N\Xi}$ can be determined by imposing the $\Xi$-potential in uniform matter $v_{\Xi}^{unif}=\partial\epsilon/\partial\rho_\Xi$ to be equal to the empirical value $U_\Xi$ at saturation density in the absence of $\Lambda$ and $\Xi$, leading to:
\begin{eqnarray}
\alpha^{N\Xi} &=& -U_\Xi/\rho_0 
\end{eqnarray}
A value of $U_\Xi\approx 14$~MeV is deduced from the analysis of the spectrum of the $(K^-,K^+)$ reaction on a $^{12}$C target to produce $^{12}_\Xi$Be, assuming
a Woods-Saxon potential for the $\Xi^-$ potential~\cite{Khaustov2000}. This yields $\alpha^{N\Xi} \approx$ 100~\hbox{MeV fm$^3$}.
Another, and maybe more direct way, to determine the parameter $\alpha^{N\Xi}$ is to calculate the $\Xi^-$ removal energy $B_{\Xi^-}$ defined as,
\begin{eqnarray}
B_{\Xi^-} = E_{tot}(N,Z)-E_{tot}(N,Z,N_{\Xi^-}),
\label{eq:kiso}
\end{eqnarray}
where $E_{tot}$ is  { the total binding energy, that is} the total energy with subtraction of the rest mass term~ \cite{Sun2016}.
This allows to compare to two experimental energies for $^{12}_{\Xi s}$Be~\cite{Khaustov2000} (with N=6, Z=5) and 
$^{15}_{\Xi }$C~\cite{Nak2015} (with N=7, Z=7), also called the "Kiso" event.
The latter experimental data is however subject to two possible interpretations: 1) assuming that $^{15}_{\Xi}$C is produced in its ground-state ($\Xi$ being in the 1s single particle state), or 2) assuming that $^{15}_{\Xi}$C is produced in its first excited-state ($\Xi$ being in the 2p single particle state). A recent theoretical analysis based on mean-field theory has shown that interpretation 2) is also compatible with the removal energy of $^{12}_{\Xi s}$Be~\cite{Sun2016}.
We have performed a similar analysis with our density-functional as illustrated in Tab.~\ref{table:removal}. 
In our model, the $\Xi^-$ removal energies $B_{\Xi^-}$ is uniquely determined by the parameter $\alpha_1^{N\Xi}$, which we vary around 100~MeV~fm$^{3}$.
It is shown in Tab.~\ref{table:removal} that the value $\alpha^{N\Xi}=109$~MeV~fm$^{3}$ which reproduces well the $\Xi^-$ removal energies supposing $^{15}_{\Xi}$C in its ground-state is not compatible with the expected $\Xi^-$ removal energies of $^{12}_{\Xi s}$Be, while the value $\alpha^{N\Xi}=125$~MeV~fm$^{3}$ which reproduces well the $\Xi^-$ removal energies supposing $^{15}_{\Xi}$C in its first excited-state gives reasonable results for the expected $\Xi^-$ removal energies of $^{12}_{\Xi s}$Be and 
$^{15}_{\Xi s}$C~\cite{Sun2016}.
In the following, we thus fix $\alpha_1^{N\Xi}=125$~MeV~fm$^{3}$.

In the case of the hyperon-hyperon couplings, since it is yet impossible to fix the values of these parameters from experimental data, they
are normalized to better known parameters such as $\alpha^{N\Lambda}$, $\alpha^{N\Sigma}$
$\alpha^{N\Xi}$ and $\alpha^{\Lambda\Lambda}$.The following dimensionless parameter are therefore introduced:
\begin{equation}
\beta^{YY^\prime}=\frac{\alpha^{YY^\prime}}{\alpha^{NY^\prime}}, \hspace{0.5cm} Y\neq Y^\prime
\end{equation}
with $Y=\Xi$ and $Y^\prime=\Lambda$ and $\Xi$,
\begin{equation}
\beta^{YY}=\frac{\alpha^{YY}}{\alpha^{\Lambda\Lambda}}.
\end{equation}

The parameters $\beta^{YY^\prime}$ and $\beta^{YY}$ are largely unknown.
They are expected to be less influential than the parameters in the $NY^\prime$ and $\Lambda\Lambda$ channels since they act between minority species.
The channels of interest in the present study are $\Lambda\Xi$ and $\Xi\Xi$.
For the same argument related to the number of particles, we expect i) these channels to be rather weak, and ii) that the $\Lambda\Xi$ channel is more
influential than the $\Xi\Xi$ channel .

\begin{table}[t]
\renewcommand{\arraystretch}{1.3}
\begin{ruledtabular}
\begin{tabular}{cccccccccc}
                                              & VY0 & VY1 & VY2 \\
\hline
 $\alpha_1^{N\Xi}$                  & 125  & 125 & 125 \\
 $\beta^{\Lambda\Xi}$            &  0      & 1 & 1 \\
 $\beta^{\Xi\Xi}$                      & 0      & 0 &  1 \\
\end{tabular}
\end{ruledtabular}
\caption{Set of values for the dimensionless parameters $\alpha_1^{NY}$ and $\beta^{YY^\prime}$ used in this work. }
\label{table:VSX}
\end{table}

Since the interaction in the $\Lambda\Xi$ and $\Xi\Xi$ channels is unknown, it is difficult to do more than a sensitivity analysis.
We have defined three models, VY0-2, for the sensitivity analysis which are given in table~\ref{table:VSX}. 
The sign of the interaction parameters is not relevant for the sensitivity analysis, and it is arbitrarily chosen positive.
The influence of these choices will be studied in Section~\ref{sec:groundstates}.

The mean field potentials in uniform matter $v_\Lambda^{unif}=\partial\epsilon/\partial\rho_\Lambda$,  and
$v_\Xi^{unif}=\partial\epsilon/\partial\rho_\Xi$ are displayed in Fig.~\ref{fig:vt} as a function of the 
nucleon density $\rho_N$ (in units of the saturation density $\rho_0$) for the functional DF-NSC89+EmpC+VY0
and for various choices of the densities $\rho_\Lambda$ and $\rho_\Xi$ expressed in fm$^{-3}$ (a),
and for various strength of the $\Lambda\Xi$ interaction (b).

Let us first discuss the potential in uniform matter $v_\Lambda^{unif}(\rho_N,\rho_\Lambda,\rho_\Xi)$ shown in
Fig.~\ref{fig:vt}(a).
The addition of a finite amount of $\Lambda$ for $\rho_\Lambda$=0.03~fm$^{-3}$ to standard nuclear matter, increases the value of the
potential (by about 5~MeV at $\rho_0$), while the addition of the same amount of
$\Xi$ decreases the potential (by about -5~MeV at $\rho_0$).
The larger the $\Lambda N$ repulsion ($\Lambda\Xi$ attraction) the shallower (deeper) the potential.
It should also be noted that the very same attractive term in the energy functional $\epsilon_{\Lambda\Xi}$ is responsible
for the decrease of $v_\Lambda^{unif}(\rho_N,\rho_\Lambda,\rho_\Xi)$  and
the decrease of $v_\Xi^{unif}(\rho_N,\rho_\Lambda)$ by adding $\Xi$. 
 If it was not for the cost in rest mass, it would therefore be preferable to add 
$\Xi$-hyperons than $\Lambda$-hyperons, due to the gain in the $\Lambda$ potential  {, as well as the reduced Pauli blocking of $\Xi$ single-particle states in $N\Lambda$ matter}.
If instead of being attractive, the $\Lambda\Xi$ channel is repulsive, the effect at the level of the potential
would be opposite to the present case.

\begin{figure}[tb]
\begin{center}
\includegraphics[width=0.5\textwidth]{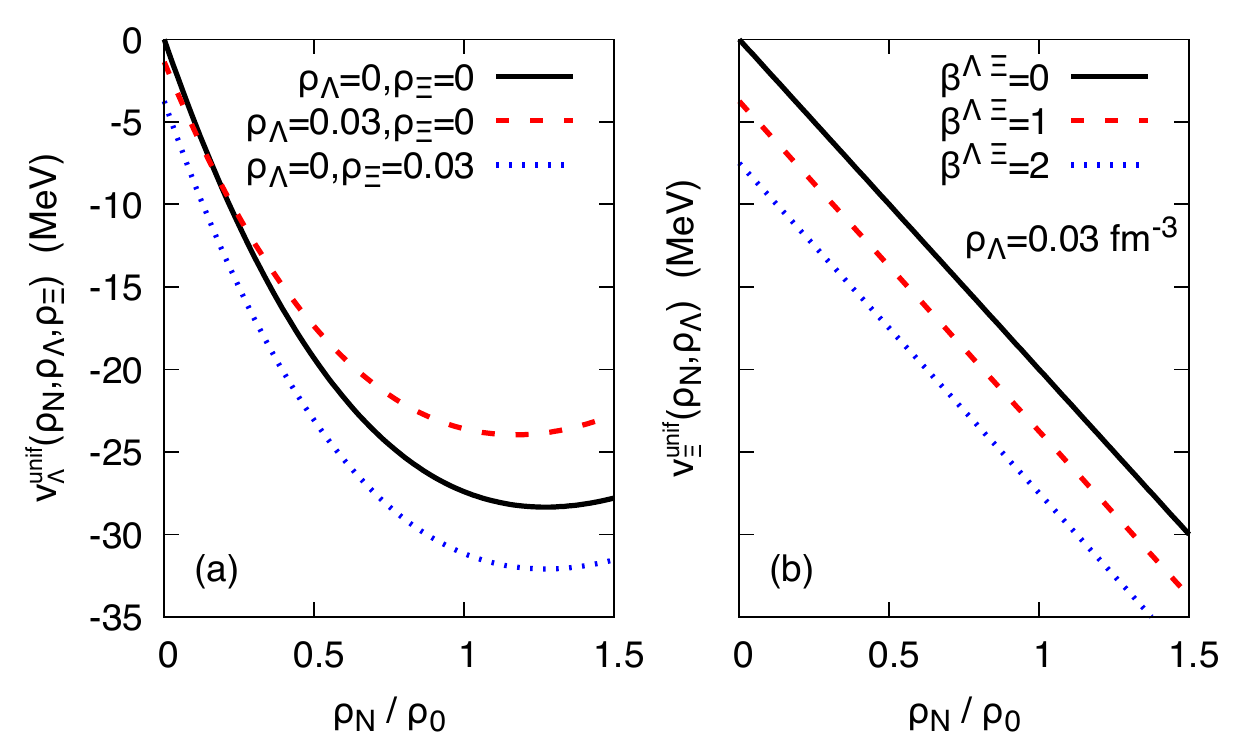}
\caption{(Color online) Hyperon potentials in uniform matter $v_\Lambda^{unif}(\rho_N,\rho_\Lambda,\rho_\Xi)$ (a) and
$v_\Xi^{unif}(\rho_N,\rho_\Lambda)$ (b) as a function of the nucleon density $\rho_N$ 
(in units of the saturation density $\rho_0$) for the functional DF-NSC89+EmpC+VY0
for various choices of the densities $\rho_\Lambda$ and $\rho_\Xi$ (in fm$^{-3}$) (panel a),
and for various strength of the $\Lambda\Xi$ interaction (panel b). See text for more details.}
\label{fig:vt}
\end{center}
\end{figure}

The potential $v_\Xi^{unif}(\rho_N,\rho_\Lambda)$ is displayed in Fig.~\ref{fig:vt}(b)
for various choices of the strength of the $\Lambda\Xi$ interaction (represented by the parameter $\beta^{\Lambda\Xi}$)
and for a fixed amount of $\Lambda$. 
This figure shows that the larger is the parameter $\beta^{\Lambda\Xi}$, the more attractive is the potential $v_\Xi^{unif}(\rho_N,\rho_\Lambda)$, as expected.

\section{The extended hyper-nuclear chart}
\label{sec:groundstates}

\begin{table}[t]
\caption{ Parameters of the $\Lambda$-effective mass given by Eq.~(\ref{mfit}) for the functionals considered in this work.}
\begin{ruledtabular}
\begin{tabular}{l|ccccccccccccccccc}
Force & $\mu_1^{N\Lambda}$ & $\mu_2^{N\Lambda}$ & $\mu_3^{N\Lambda}$ & $\mu_4^{N\Lambda}$ \\
& & fm$^3$ & fm$^6$ & fm$^9$ \\
\hline
DF-NSC89~\cite{Cugnon2000,Vidana2001}  & 1      & 1.83 & 5.33 & 6.07 \\
DF-NSC97a~\cite{Vidana2001} & 0.98 & 1.72 & 3.18 & 0 \\
DF-NSC97f~\cite{Vidana2001} & 0.93 & 2.19 & 3.89 & 0  \\
\end{tabular}
\end{ruledtabular}
\label{table:SZ2}
\end{table}

Minimizing the total energy defined from the density functional~(\ref{functionaltot}), and using the 
Skyrme model for the nucleonic part~\cite{ben03}, we obtain the usual Schr\"odinger equation ($i=N,Y$),
\begin{eqnarray}
&&\Big[-\nabla\cdot \frac{\hbar^2}{2 m^*_i(r)}\nabla+V_i(r)-iW_i(r)(\nabla\times\sigma)\Big]\varphi_{i,\alpha}(r)
\nonumber \\
&&\hspace{4cm}=-e_{i,\alpha}\varphi_{i,\alpha}(r), \label{eq:schrod}
\end{eqnarray}
where $W_i$ is the spin-orbit potential~\cite{Book:RingSchuck} and the nucleon potential $V_N$ is defined as,
\begin{eqnarray}
V_{N}(r)&=&v_{N}^{Skyrme}+ \frac{\partial}{\partial \rho_N}\left( \frac{m_\Lambda}{m_\Lambda^*(\rho_N)} \right) \times \nonumber \\
&&\hspace{-1cm} \left( \frac{\tau_\Lambda}{2m_\Lambda}-\frac{3}{5}\frac{(3\pi^2)^{2/3}\hbar^2}{2m_\Lambda}\rho_\Lambda^{5/3}
\right) ,
\label{eq:VN}
\end{eqnarray}
The $\Lambda$-hyperon potential $V_\Lambda$ is given by,
\begin{eqnarray}
V_{\Lambda}(r)&=&v_{\Lambda}^{unif}
-\left( \frac{m_\Lambda}{m_\Lambda^*(\rho_N)} -1\right)
\frac{(3\pi^2)^{2/3}\hbar^2}{2m_\Lambda}\rho_\Lambda^{2/3} .
\label{eq:VL}
\end{eqnarray}
and the $\Lambda$ effective mass $m_\Lambda^*$ determined from BHF calculations~\cite{Cugnon2000,Vidana2001} is expressed as
\begin{equation}
\frac{m_\Lambda^*(\rho_N)}{m_\Lambda} = \mu_1^{N\Lambda}-\mu_2^{N\Lambda}\rho_N+\mu_3^{N\Lambda}\rho_N^2-\mu_4^{N\Lambda}\rho_N^3 .
\label{mfit}
\end{equation}
The values for the parameters $\mu_{1-4}$ for the functional considered here are given in Table~\ref{table:SZ2}.

For the $\Xi$ hyperon potentials, we have the following relation
\begin{eqnarray}
V_\Xi(r) &=& v_\Xi^{unif}=\partial\epsilon/\partial\rho_\Xi  .
\end{eqnarray}

It should be noted that in the case of $\Xi^-$, an additional contribution to the Coulomb potential shall be considered.
The Coulomb energy is generated by the Coulomb interaction among charged particles p and $\Xi^-$. It is decomposed into a direct term,
\begin{eqnarray}
E_\mathrm{Coul}^D &=& \sum_{i\ne j}
\mathrm{sgn}(i) \mathrm{sgn}(j) \nonumber \\
&&\frac{e^2}{2} \int d^3\mathbf{r} d^3\mathbf{r'} \rho_i(r) \frac{1}{|\mathbf{r}-\mathbf{r'}|}\rho_j(r') ,
\label{app:ecould}
\end{eqnarray}
where $i,j=p,\Xi^-$ and $\mathrm{sgn}(i)$ is the sign of the Coulomb charge of particle $i$.
It should be noted that the $p\Xi^-$ channel is attractive with respect to the Coulomb direct interaction, which could favour the onset of the $\Xi^-$ hyperon against $\Xi^0$.

Considering the Slater approximation, the exchange term reads
\begin{eqnarray}
E_\mathrm{Coul}^E &=& - e^2 \frac 3  4 \left( \frac{3}{\pi}\right)^{1/3} \int d^3\mathbf{r}  \left( \rho_p^{4/3} + \rho_{\Xi^-}^{4/3} \right) \, .
\label{app:ecoule}
\end{eqnarray}
The exchange term is attractive for all charged particles, favouring again $\Xi^-$ against $\Xi^0$.

The contribution of the Coulomb interaction to the mean fields are obtained
by functional derivation of Eqs.~(\ref{app:ecould}) and (\ref{app:ecoule}), giving from the proton direct term,
\begin{eqnarray}
u^D_\mathrm{Coul, p}(r) &=& e^2 \int d^3\mathbf{r'}  \frac{1}{|\mathbf{r}-\mathbf{r'}|}\rho_{ch}(r') .
\end{eqnarray}
where the charge density is $\rho_{ch}= \rho_p-\rho_{\Xi^-}$.
It should be noted that the direct Coulomb terms for all other particules are defined exactly the 
same as for the proton case, with only a sign difference which refers to the charge of the considered hyperon. 
\begin{eqnarray}
u^D_\mathrm{Coul, {\Xi^-}}(r) &=&   - u^D_\mathrm{Coul, p}(r) \, .
\label{eq:coulDY}
\end{eqnarray}
We consider the extension of the Slater approximation for multi-types of charges particles, giving for the exchange Coulomb potential
\begin{eqnarray}
u^E_\mathrm{Coul, i}(r) &=& - e^2  \left( \frac{3}{\pi}\right)^{1/3} 
 \int d^3\mathbf{r'} \left\{\rho_i(r')\right\}^{1/3} \, ,
\end{eqnarray}
where $i=p$, ${\Xi^-}$.
As we noticed, the Coulomb interaction favour the onset of negatively charged particles over neutral ones.
The $\Xi^-$ hyperon could therefore be favoured against $\Xi^0$.

\subsection{Numerical strategy}
\label{sec:numerics}

The HF equations are solved in coordinate representation assuming spherical symmetry.
Deformations are known to play an important role in the structure of light hypernuclei~\cite{Zhou2007,ThiWin2008,Schulze2010}, however this approximation is expected to hold at the level of   accuracy in our work. Indeed deformation induces corrections to energies which approximately scale as $A^{-1/6}$, and can be neglected when calculating energy differences of nuclei with $A>20$ as it is done in this work for the calculation of the $\Xi$-instability phenomenon.

To correct for the spurious one-body center of mass energy,  the mass $m_i$ of each species $i$ in the Schr\"odinger equation is replaced by the reduced mass $m^\prime_i$, defined as
$(m^\prime_i)^{-1} = {m^{-1}_i} - (\sum_{j\ne i} N_j m_j)^{-1}$, where $i$ and $j$ indexes run over $N$ and $Y=\Lambda$, $\Xi^{0,-}$. 

The Numerov method is used to determine the wave-functions $\varphi_{i,\alpha}(r)$ for given potentials $V_i(r)$ and $W_i(r)$ as well as given effective mass
$m^*_i(r)$ and we consider the vanishing wave-function Dirichlet boundary condition. 
The coordinate space extends up to 30~fm and it is discretised with equal steps of 0.1~fm.
Masses of particles are fixed to be their bare masses, except for neutrons, protons and $\Lambda$ which acquire an effective mass in dense medium.
As usual in HF solvers, the self-consistency is reached by successive iterations until the total energy converges within an accuracy of less than $10^{-8}$~MeV.
Further details about the implementation of the Hartree-Fock approach in the hypernuclear case can be found in Refs.~\cite{Cugnon2000,Vidana2001,Khan2015} for instance.

When considering nucleons and hyperons (here $\Lambda$ and $\Xi^{0,-}$), the three conserved charges of the strong interaction are defined as:
\begin{eqnarray}
A&=&N_n+N_p+N_Y,\\
Q&=&N_p-N_{\Xi^-},\\
-S&=&N_\Lambda+2N_{\Xi},
\end{eqnarray}
where the hyperon numbers are
\begin{eqnarray}
N_Y&=&N_\Lambda+N_\Xi,\\
N_\Xi&=&N_{\Xi^+}+N_{\Xi^0}
\end{eqnarray}

In the following, hyperons are added on top of a core-nucleus ($A_{core}$, $Z_{core}$) with step in strangeness $\Delta S=2$.
The S-drip line (S-DL) is the drip line in the strangeness number $S$. This is the maximum value for -S before the chemical potential of any of the hyperon becomes positive.

As discussed in the introduction, according to the free space Q-values, the hypernuclear ground state is expected to contain only $\Lambda$'s for low strangeness $-S$, followed by $\Xi^{0,-}$ when the number of $\Lambda$'s is sufficiently high for the gain in kinetic energy to compensate the energy cost in rest mass. 
The  $\Xi^{0,-}$ instability is defined as the strangeness number $-S$ at which the first $\Xi^{0,-}$ appear in the hypernucleus ground state.
This number is called in the following S$_\mathrm{inst.}$.
In the present work we limit the exploration in $S$ up to S$_\mathrm{inst.}$.
The appearance of  $\Xi^{0,-}$ is given from the comparison of the energies of the $\Lambda$-hypernucleus to all other ground states formed by $\Lambda$ and $\Xi^{0,-}$ with same mass A, total charge Q (or proton charge $Z$, in the case where constant $Z$ transformations are considered), and strangeness S numbers.
Especially, we compare the energy of the system  $A_{core}+N_{\Lambda}+N_{\Xi}$ to the energy of all the systems composed of $(A_{core}+1)+(N_{\Lambda}-2)+(N_{\Xi}+1)$ , which correspond to the transformation of 2$\Lambda$ into one $\Xi$ and a $N$.

\begin{figure}[tb]
\begin{center}
\includegraphics[width=0.48\textwidth]{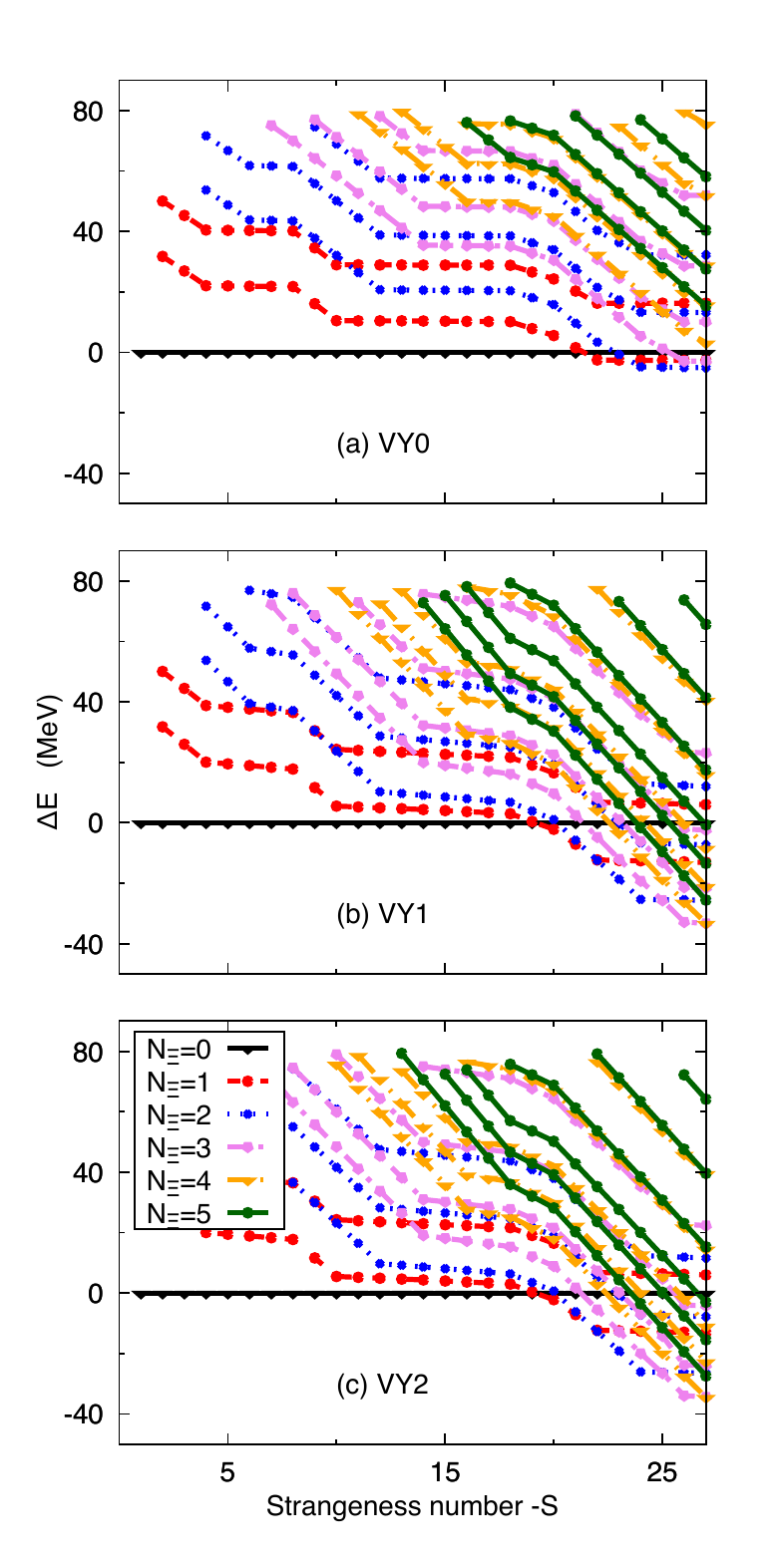}
\caption{(Color online) Representation of the energy difference $\Delta E_{tot}(S,N_\Xi)=E_{tot}(N_\Lambda=-S-N_\Xi/2,N_\Xi)-E_{tot}(N_\Lambda=-S,N_\Xi=0)$
for various multi-strange hypernuclei conserving the mass A=132, the electric charge Q=50, and for which the strangeness charge -S is varied from 0
to 30, for the functionals SLy4+DF-NSC89+EmpC+VY0 (a), VY1 (b) and VY2 (c).}
\label{fig:sn1}
\end{center}
\end{figure}

An illustration of the search for the $\Xi^{0,-}$ instability is shown in Fig.~\ref{fig:sn1}, for the case $A=132$, $Q=50$. 
The different colors correspond to different number of $\Xi$ as indicated in the box. 
 Up to the strangeness number $-S\approx 20$ the configuration with only $\Lambda$'s corresponds to the hypernucleus ground state. 
The energy difference between the configurations with a given number of $\Xi$ is due i) to the slight mass difference between $\Xi^0$ and $\Xi^-$, ii) to the Coulomb energy and iii) to the Pauli blocking effect.
In this case, for same $N_\Xi$ groups the lowest energy configurations are always the ones with the largest number of $\Xi^{-}$.
The sharp energy drops reflect shell closures. 
Considering the lowest energy configuration for each $N_\Xi$ groups, the energy hierarchy scales well with $N_\Xi$ up to $S\approx -20$ but at larger values of -$S$, the different configurations are highly degenerated, and the composition of the actual ground state shall depend on the hypothesis for the unknown couplings. 
These unknown coupling are varied from VY0 to VY2, see Table~\ref{table:VSX}.
The $\Xi^{0,-}$ instability is only weakly impacted by the choice for the unknown couplings, while the ground-state energy beyond the $\Xi^{0,-}$ instability is largely impacted.
The largest uncertainty comes from the $\Lambda\Xi$ interaction (VY1), while the $\Xi\Xi$ interaction (VY2) seams to be weakly influential, even for a finite amount of $\Xi$.
It is not surprising that the largest impact comes from the $N\Xi$ channel (VY0) since $N$ is the dominant species, then comes the $\Lambda\Xi$ channel (VY1) and finally the weakest channel is the $\Xi\Xi$ (VY2) one. For this same reason, the influence of the repulsive Coulomb interaction 
between $\Xi^-$ turns out to be very small.

In the absence of $\Lambda\Xi$ interaction (case VY0) it should be noted that the configurations with a same number of $\Xi^-$'s are almost degenerate beyond S=S$_\mathrm{inst.}$. The introduction of $\Lambda\Xi$ interaction breaks this quasi-degeneracy.

\subsection{The $\Xi^{0,-}$ instability over the nuclear chart}
\label{sec:decay}

We now turn to a systematic exploration of the $\Xi^{0,-}$ instability over the nuclear chart and compare the predictions of the different functionals.
 
As already discussed in the introduction, 
the onset of $\Xi^0$ is slightly favoured over $\Xi^-$ from the mass differences.
However in dense matter, opposite effects coming from the kinetic energy term and the Coulomb interaction may play an important role.
The final results of these contradictory tendencies reflect in the chemical potentials.
We thus define the following quantities: $\Delta\mu(\Xi^{-})\equiv\mu_{\Xi^-}+\mu_p-2\mu_\Lambda$ and $\Delta\mu(\Xi^{0})\equiv\mu_{\Xi^0}+\mu_n-2\mu_\Lambda$, where the chemical potentials are defined without rest mass.

\begin{figure}[tb]
\begin{center}
\includegraphics[width=0.5\textwidth]{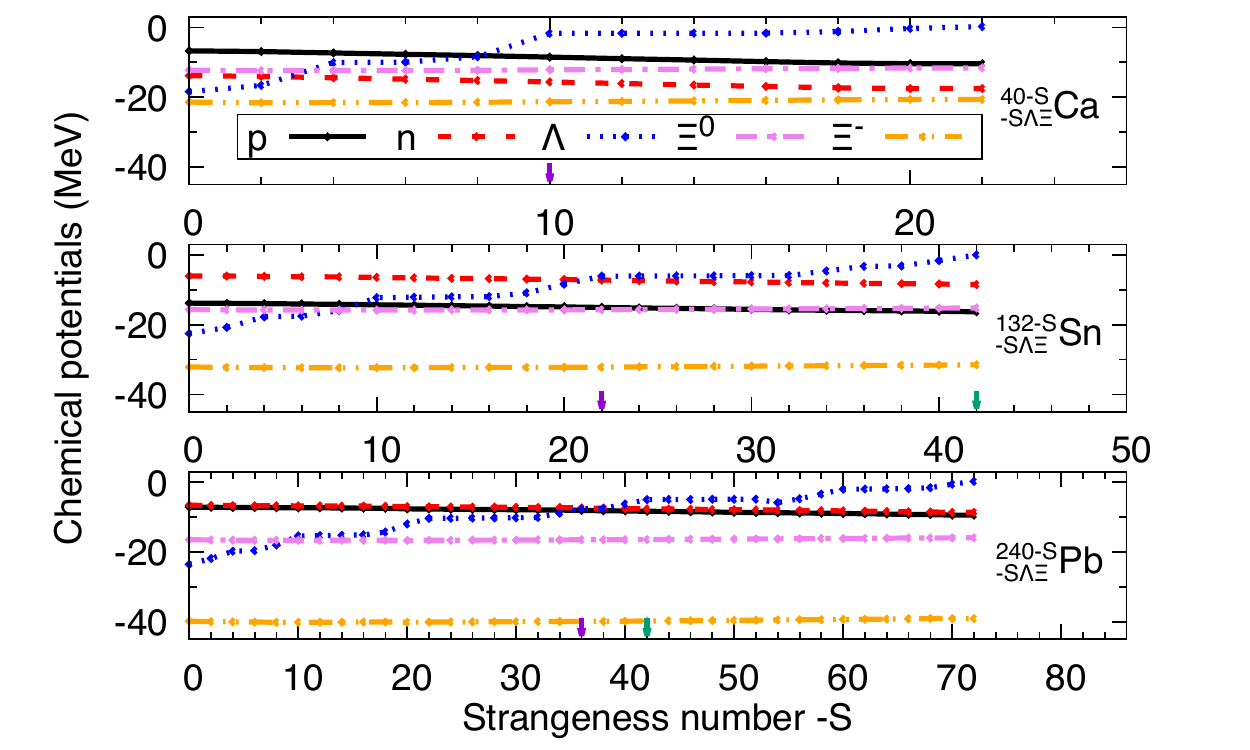}
\caption{(Color online) Chemical potential for the functionals SLY4+DF-NSC89+EmpC+VY0. }
\label{fig:chempot}
\end{center}
\end{figure}

In order to evaluate the mean field contribution to the onset properties, the evolution of the chemical potentials is displayed in Fig.~\ref{fig:chempot}
up to the S-DL for pure-$\Lambda$ hypernuclei.
The following typical core nuclei are considered: $^{40}$Ca, $^{132}$Sn, and $^{208}$Pb, on top of which strangeness is added.
The position of the S$_\mathrm{inst.}$ is indicated by the vertical arrows for conserved $Q$ (purple arrow) or conserved $Z$ (green arrow).

As expected from the Coulomb interaction, the chemical potential of the $\Xi^-$ is lower than that of the $\Xi^0$ for all hypernuclei. 
The difference between these chemical potentials is already of about 5~MeV for the lightest system shown in Fig.~\ref{fig:chempot},
and reaches about 40 MeV for the heaviest nuclei.
This observation confirms the specific role played by the $\Xi^-$, especially for systems studied at conserved $Q$.
It highlights the contribution of the Coulomb field in the correction to the mean field for negatively charged particles.
We can therefore anticipate that $\Xi^-$ will certainly appear as the first particle in most of the cases.

Moreover, it should be noted that the $\Xi$-instability at conserved $Q$ occurs when the $\Lambda$ chemical potential crosses the neutron or proton chemical potential.
At conserved $Z$, the $\Xi$-instability is observed for larger values of the strangeness number -S.
There is no $\Xi$-instability at conserved $Z$ in Ca, and as the mass increases, the $\Xi$-instability at conserved $Z$ comes closer to that at conserved $Q$.
The reason is because in light systems, there is a big gap between the onset of the $\Xi^-$ (first appear at conserved $Q$) and of the $\Xi^0$ (single system allowed at conserved $Z$).
This energy gap tends to become less and less important as the charge of the system increases.

\begin{figure}[tb]
\begin{center}
\includegraphics[width=0.5\textwidth]{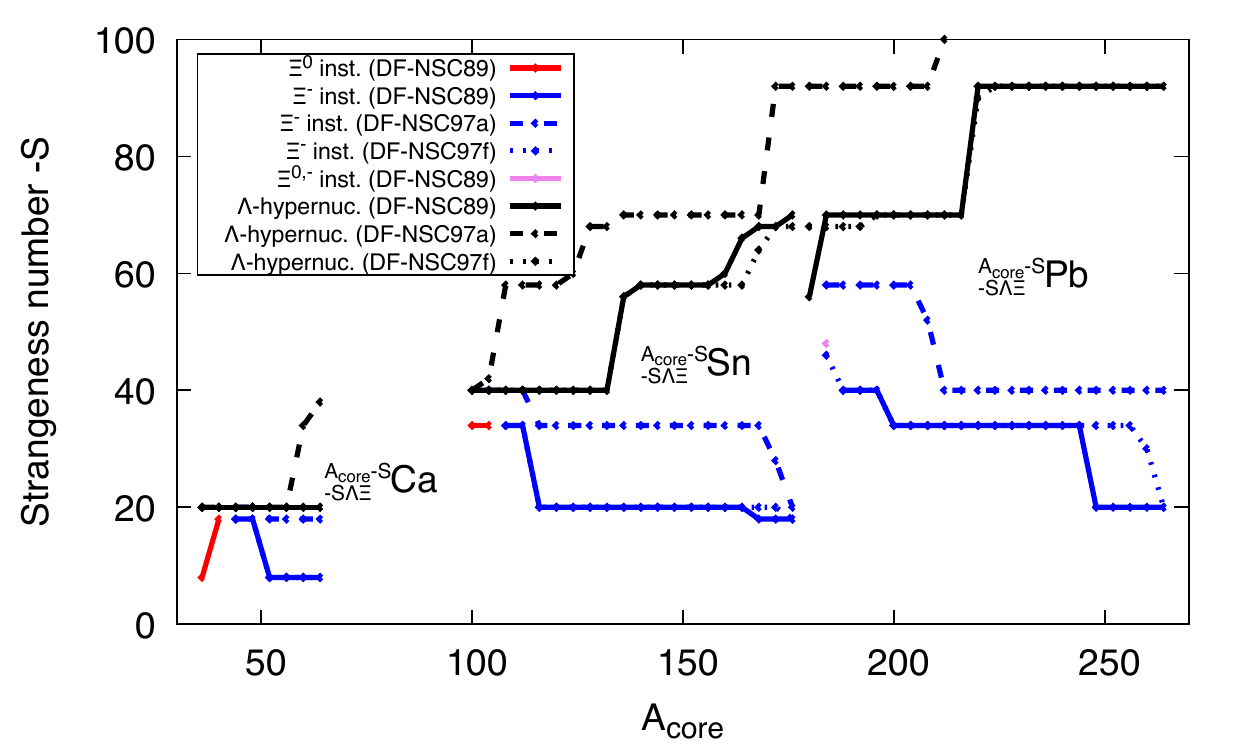}
\caption{(Color online) Comparison of the strangeness number $-S$ at the S-drip line for $\Lambda$-hypernuclei (black lines) with S$_\mathrm{inst.}$, the strangeness number $-S$ at the $\Xi^{0,-}$ instability (red, blue and pink line-points).
A sample of various total charge $Q$ is considered, $Q=20$, $50$, and $82$, and $A_{core}$ runs from the proton to the neutron drip line.. Results from functional SLy4+DF-NSC89+EmpC+VY0.}
\label{fig:sdll-all}
\end{center}
\end{figure}

\begin{figure*}[b]
\begin{center}
\includegraphics[width=0.9\textwidth]{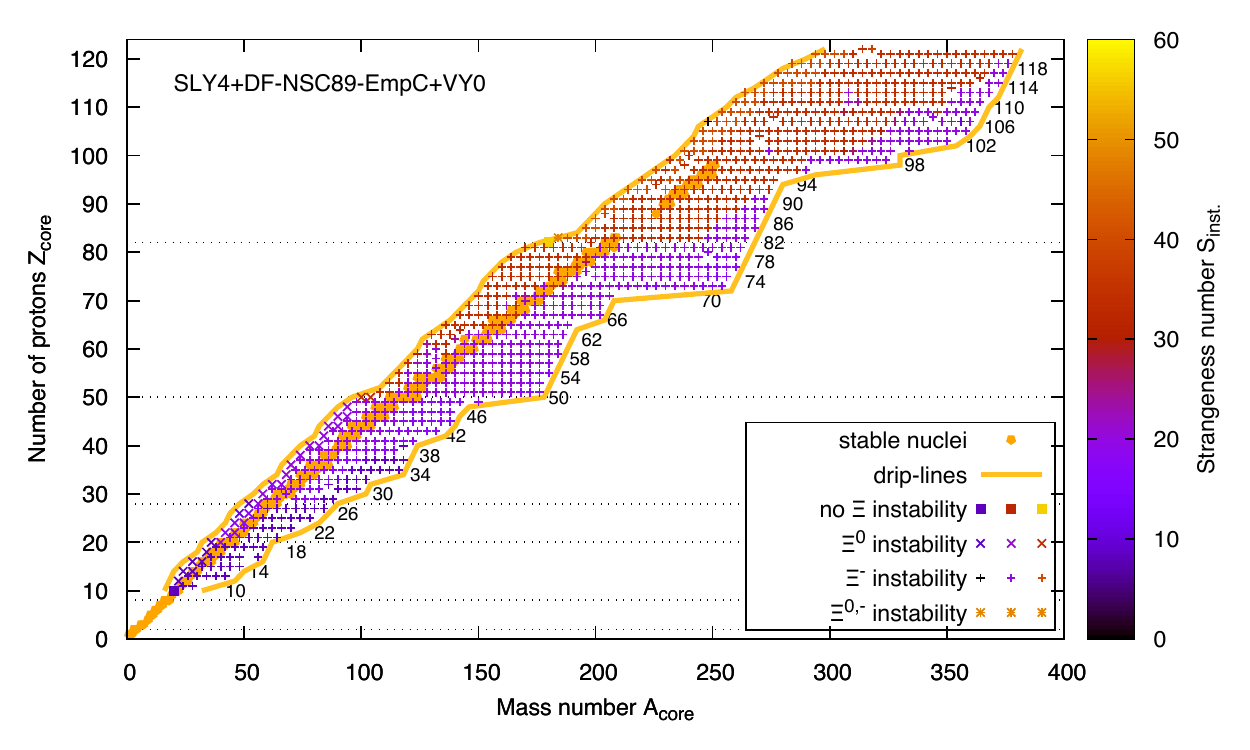}
\caption{(Color online) Prediction for  S$_\mathrm{inst.}$ through the nuclear chart. Calculations are performed with
the density functional DF-NSC89+EmpC for the $\Lambda$ interaction, and SLy4 and VY0 for the nucleon and other hyperons interactions.
Each point represents a calculation performed at constant $A$ and $Q$ (total charge) and they are represented as function of the baryonic number $A_{core}$ and charge $Z_{core}$ associated to neutrons and protons.
The value of the total charge $Q$ is written at the end of each iso-charge lines. See text for more details.}
\label{fig:chart89C}
\end{center}
\end{figure*}

\begin{figure*}[tb]
\begin{center}
\includegraphics[width=0.9\textwidth]{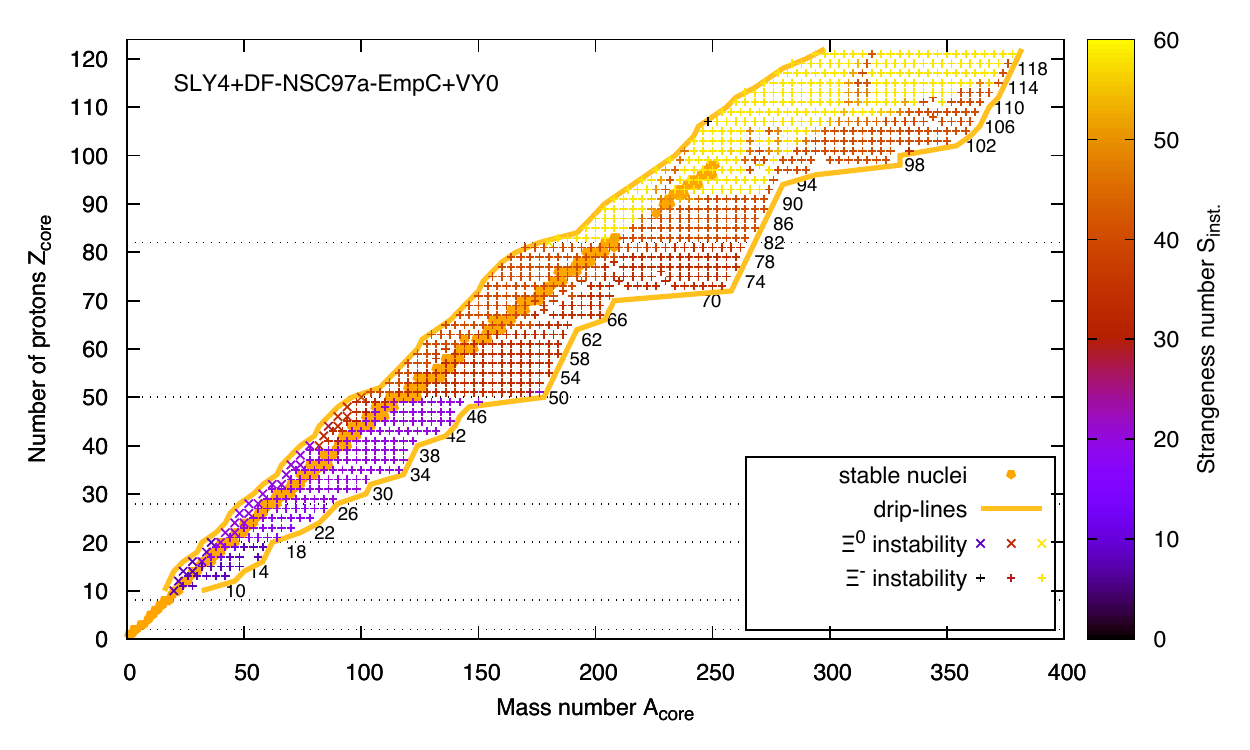}
\caption{(Color online) Same as Fig.~\ref{fig:chart89C} using the functional DF-NSC97a+EmpC}.
\label{fig:chart97aC}
\end{center}
\end{figure*}

We now come to more systematics by analyzing Ca, Sn and Pb isotopes.
Fig.~\ref{fig:sdll-all} displays a comparison of the S-drip line obtained for $\Lambda$-hypernuclei (black lines), as obtained in Ref.~\cite{Khan2015}, with the value of S$_\mathrm{inst.}$ associated to the onset of the first $\Xi^{0}$ and $\Xi^{-}$ hyperons (red and blue lines).
The onset of the first $\Xi^{0,-}$ hyperon occurs before the S-DL for pure $\Lambda$ multi-strange hypernuclei is reached, for all hypernuclei shown in Fig.~\ref{fig:sdll-all} except one, 
namely  $^{236}_{56\Lambda}$Pb.
In this exceptional case the lowest energy state at the $\Xi$ onset is composed of 1$\Xi^0$ and 1$\Xi^-$.
The concurrent hypernucleus composed of $\Xi^-$ has a lower energy, but its proton chemical potential is about 2~MeV higher.
 Since this nucleus is close to the proton drip line, this increase makes it proton-unstable.
It should be noted that the next calculated nucleus of this isotopic chain, $^{236}_{48\Lambda,2\Xi}$Pb, also exhibits a configuration with 1$\Xi^0$ and 1$\Xi^-$ at the $\Xi$-instability.
 
At constant $Q$, the presence of $\Xi^-$ implies an extra proton, which explains why 
for extreme neutron deficient hypernuclei at the proton drip-line, 
 $\Xi^0$ are favoured at the $\Xi^{0,-}$ instability.
 
The observed plateaus are due to
strong shell effects for both the $\Xi^{0,-}$ instability and the S-DL for pure $\Lambda$-hypernuclei.
DF-NSC89 and DF-NSC97f predict similar results, while DF-NSC97a pushes up both the $\Xi^{0,-}$ instability and the S-DL for pure $\Lambda$-hypernuclei.
This is in agreement with the fact that DF-NSC97a predict $\Lambda$-matter more stable than the other functionals.
The sensitivity of S$_\mathrm{inst.}$ on the $N\Lambda$ interaction is quite large, 
but the global trend is an increasing value for -S$_\mathrm{inst.}$ with increasing nuclear mass.

We can also compare to the other estimation of S$_{inst.}$~\cite{Balberg1994} based on RMF Lagrangians.
The data used in Ref.~\cite{Balberg1994} to calibrate the model are roughly the same as ours, except for the $\Lambda\Lambda$ channel which was considered  more attractive: 5~MeV versus about 1~MeV now.
For a core of $^{208}$Pb, they have estimated -S$_{inst.}=41$ while we predict -S$_{inst.}=$36-40 depending on the $N\Lambda$ interaction.
The fact that the prediction in Ref.~\cite{Balberg1994} is slightly higher than our is mostly explained by the different choice made for the  $\Lambda\Lambda$ interaction.

\begin{figure*}[tb]
\begin{center}
\includegraphics[width=0.9\textwidth]{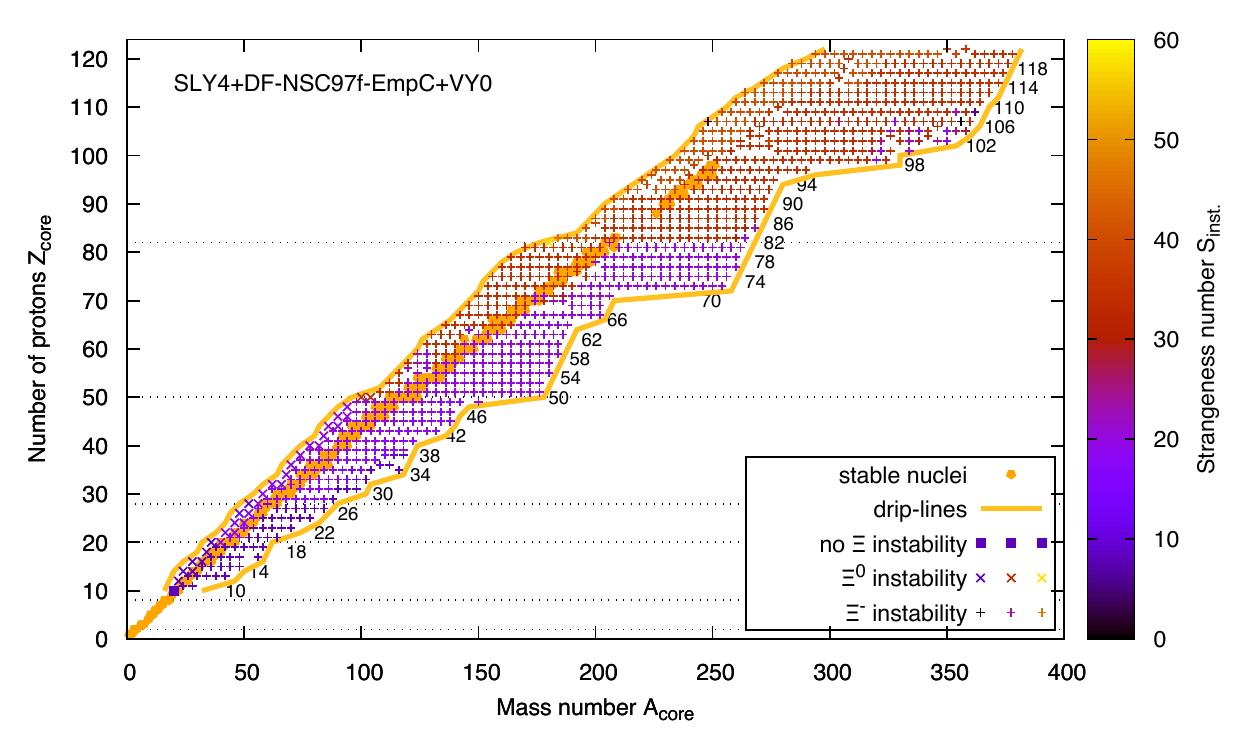}
\caption{(Color online) Same as Fig.~\ref{fig:chart89C} using the functional DF-NSC97f+EmpC}.
\label{fig:chart97fC}
\end{center}
\end{figure*}

Extended predictions for
 S$_\mathrm{inst.}$ over the nuclear chart are 
displayed in Figs.~\ref{fig:chart89C}-\ref{fig:chart97fC}. 
We explore the nuclear chart delimited by the neutron and proton drip-lines, which are defined for ordinary nuclei (with only neutrons and protons).

Calculations for $Z_{core}$ between 10 and 120 are performed, with steps 
  $\Delta A_{core}=4$ , $\Delta Z_{core}=2$, and 
$\Delta S=2$.
As discussed above, the maximum number of strangeness S$_\mathrm{inst.}$ is reached in two cases: either the S-DL is reached before the onset of the $\Xi^{0,-}$ instability (it occurs only for a few cases around $Z_{core}=82$ and $A_{core}=182$ in Figs.~\ref{fig:chart89C}-\ref{fig:chart97fC}), or the $\Xi^{0,-}$ instability is reached before the S-DL and we indicate in Figs.~\ref{fig:chart89C}-\ref{fig:chart97fC} if the first $\Xi$ to appear is a $\Xi^0$ (with x symbol) or $\Xi^-$ (with + symbol).

In Figs.~\ref{fig:chart97aC} and \ref{fig:chart97fC}, we compare the predictions for the $\Xi^{0,-}$ instability considering the functionals DF-NSC97a+EmpC and DF-NSC97f+EmpC.
One of the main difference between these two functionals is that DF-NSC97a+EmpC predicts a higher strangeness at the $\Xi^{0,-}$ instability -$S_{inst.}$ than DF-NSC97f+EmpC.
This feature is consistent with the fact that the $N\Lambda$ mean field predicted by DF-NSC97a+EmpC is more attractive 
, leading to more bound 
$\Lambda$-hypernuclei.

Despite some quantitative differences between the predictions for the $\Xi^{0,-}$ instability shown in 
Figs.~\ref{fig:chart89C}-\ref{fig:chart97fC}, gross features emerge from the comparison of the results obtained with various
$\Lambda$-interactions:
\begin{itemize}
\item  The instability with respect to the onset of $\Xi^{0,-}$ is observed all along the nuclear chart. Only a very small region of proton rich nuclei around $Z_{core}=82$ may not be $\Xi$-unstable (only for DF-NSC89).
\item In most cases, the first hyperon to be formed is $\Xi^-$. The onset of $\Xi^{0}$ is predicted only for nuclei close to the proton drip-line and for $Z_{core}<50$.
\item As $A_{core}$ increases the value of $S_{inst.}$ increases by steps, showing some shell effects (also visible in Fig.~\ref{fig:sdll-all}).
\end{itemize}

These predictions are very weakly influenced by the choice of the $\Lambda$ interaction.
The impact of the $\Lambda$ interaction is only observed for the absolute value of $S_{inst.}$: the softer the $\Lambda$ interaction, the higher $S_{inst.}$.
For instance, on the stability valley the softer $\Lambda$ interaction (DF-NSC97a+EmpC) predicts larger values for $S_{inst.}$ for heavy hypernuclei (up to 60) than the others
(40-50).

\subsection{Number of hypernuclei}
\label{sec:dripline}

In a previous work, we have counted the number of new multi-strange hypernuclei for system formed of N and $\Lambda$ only~\cite{Khan2015}. 
Since the $\Xi^{0,-}$ instability was not considered, we now proceed to a new counting of pure-$\Lambda$ hypernuclei up to the $\Xi^{0,-}$ instability.

\begin{table}[t]
\renewcommand{\arraystretch}{1.3}
\begin{ruledtabular}
\begin{tabular}{ccccccccccc}
            & DF-NSC89 & DF-NSC97a & DF-NSC97f \\
            & +EmpC      & +EmpC         & +EmpC     \\
$-S$        & + VY0        & + VY0        & + VY0        \\
\hline            
\multicolumn{4}{c}{Ordinary nuclei} \\
0          & 7 688 (1 922) & 7 688  (1 922)  & 7 688  (1 922)  \\
\hline            
\multicolumn{4}{c}{pure-$\Lambda$ hypernuclei below $\Xi$-instability} \\
2            & 7 664  (1 916)  & 7 656  (1 922)  & 7 664  (1 916)  \\
8            & 7 664  (1 916)  & 7 656  (1 922)  & 7 648  (1 912)  \\
20          & 6 352  (1 588)  & 7 248  (1 812)  & 6 544  (1 636)  \\
40          & 520  (130)        & 4 576  (1 144)  &    872  (218)  \\
70          & 0  (0)                & 0  (0)                & 0  (0)  \\
Total     & 198 448 (24 806) & 303 440 (37 930) & 211 744 (26 468)\\
$S_{max}$ & 56                & 58                    & 58 \\
\hline            
\multicolumn{4}{c}{pure-$\Lambda$ hypernuclei unrestricted} \\
2            & 7 672  (1 918)  & 7 664  (1 916)  & 7 688  (1 922)  \\
8            & 7 672  (1 918)  & 7 656  (1 914)  & 7 672  (1 918)  \\
20          & 7 568  (1 892)  & 7 616  (1 904)  & 7 672  (1 918)  \\
40          & 6 744  (1 686)  & 7 168  (1 792)  & 7 560  (1 890)  \\
70          & 4 576  (1 144)  & 5 896  (1 474)  & 4 344  (1 086)  \\
Total     & 604 112 (75 514) & 767 952 (95 994) & 592 192 (74 024)\\
$S_{max}$ & 140 & 180 & 140 \\
\end{tabular}
\end{ruledtabular}
\caption{Number of bound multi-$\Lambda$ hypernuclei for $10<Z_{core}<120$.
In parenthesis is indicated the number of even-even-even hypernuclei.}
\label{table:multihyp}
\end{table}

Table~\ref{table:multihyp} displays the counting of pure-$\Lambda$ hypernuclei for two cases:
first up to the $\Xi^{0,-}$ instability, and then up to strange-drip line (unrestricted).
The latter case is equivalent to our previous calculation in Ref.~\cite{Khan2015} but the counting is a bit different.
In Ref.~\cite{Khan2015} only a few strangeness numbers were considered, $-S=2,8,20,40,70$, corresponding to $\Lambda$-shell closure without spin-orbit, and
the position of the drip line was obtained by interpolation between these cases.
We found in this previous work about 490 000 $\Lambda$-hypernuclei having a maximum of 70 $\Lambda$.
In the present calculation, we systematically calculate the ground-state of hypernuclei for almost every strangeness number (step $\Delta S=2$), and
we do not limit the maximum strangeness number.
{Hypernuclei in-between shell-closure are still calculated within the spherical approximation. However, 
the effect of deformation cannot change the present estimations by more than a few percent.}
We found that the maximum strangeness number is about 56-58 below the $\Xi^{0,-}$ instability and 140-180 in the unrestricted case.
Since this maximum number is larger than the one considered in our previous work, we find a larger amount of hypernuclei in the
unrestricted case: 600 000-800 000 $\Lambda$ hypernuclei are presently predicted.
Some differences are found between the predictions of the different V$_{N\Lambda}$ functionals:
as expected, the functional DF-NSC97a predicts a larger amount of hypernuclei since the $N\Lambda$ interaction in this case is the more attractive.

All these predictions for pure-$\Lambda$ hypernuclei (unrestricted case) shall be revised since they do not consider the $\Xi^{0,-}$ instability.
Counting the number of pure-$\Lambda$ hypernuclei below the $\Xi^{0,-}$ instability, we now find that they are about 200 000 to 300 000  hypernuclei.
This number is about 1/3 to 1/2 of the unrestricted one, but it is however still very large.
It offers a considerable potential of discovery of multi-strange hypernuclei which are expected in future hypernuclear facilities.

\begin{figure}[t]
\begin{center}
\includegraphics[width=0.5\textwidth]{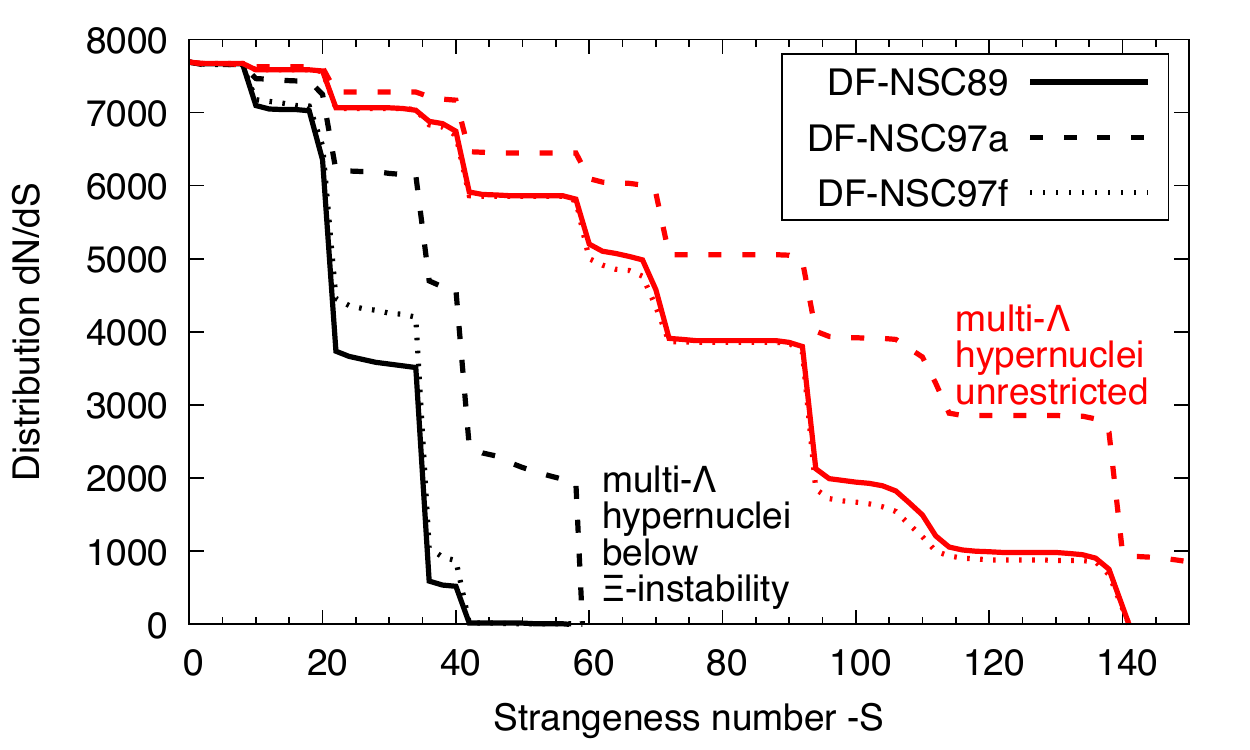}
\caption{(Color online) Distribution of hypernuclei function of the strangeness number $-S$.
The functionals that we consider are SLY4+V$_{N\Lambda}$+EmpC+VY0 where V$_{N\Lambda}$=DF-NSC89 (solid lines), DF-NSC97a (dashed lines), and 
DF-NSC97f (dotted lines). 
The black lines show the distributions of hypernuclei below the $\Xi^{0,-}$ instability while the red lines show the distribution of pure-$\Lambda$ hypernuclei (disregarding the $\Xi^{0,-}$ instability).}
\label{fig:multihyp}
\end{center}
\end{figure}

We show in Fig.~\ref{fig:multihyp} the number of pure-$\Lambda$ hypernuclei below the $\Xi^{0,-}$ instability (in black) and in the unrestricted case (in red).
The predictions for the different $N\Lambda$ functionals (DF-NSC89, DF-NSC97a, and DF-NSC97f) are shown with different line styles, see the legend in the figure.
It should be noted that the shell effects which produce the steps corresponding to shell closures (magic numbers) or sub-shell closure.
These shell closures are located at the same position for the various $N\Lambda$ functionals.
For the unrestricted case, most of the difference between DF-NSC97a and the two others is located for $-S > 70$.
This is the reason why the difference in the counting between DF-NSC97a and the two others is larger in the present case compared to our previous estimation~\cite{Khan2015}.

\begin{table}[t]
\renewcommand{\arraystretch}{1.3}
\begin{ruledtabular}
\begin{tabular}{ccccccccccc}
            & DF-NSC89 & DF-NSC97a & DF-NSC97f \\
            & +EmpC      & +EmpC         & +EmpC     \\
$-S$        & + VY0        & + VY1        & + VY2        \\
\hline            
\multicolumn{4}{c}{pure-$\Lambda$ hypernuclei below $\Xi$-instability} \\
2            & 7 664  (1 916)  & 7 664  (1 916)  & 7 664  (1 916)  \\
8            & 7 664  (1 916)  & 7 512  (1 878)  & 7 512  (1 878)  \\
20          & 6 352  (1 588)  & 4 216  (1 054)  & 4 216  (1 054)  \\
40          & 520  (130)        & 0  (0)                &    0  (0)  \\
70          & 0  (0)                & 0  (0)                & 0  (0)  \\
Total     & 198 448 (24 806) & 133 984 (16 748) & 133 776 (16 722)\\
$S_{max}$ & 56                & 34                    & 34 \\
\end{tabular}
\end{ruledtabular}
\caption{Impact of the unknown $YY$ couplings on the number of 
multi-$\Lambda$ hypernuclei below $\Xi$-instability and with $10<Z_{core}<120$.}
\label{table:multihyp2}
\end{table}

Finally, Tab.~\ref{table:multihyp2} displays a comparison for the predictions of the number of multi-$\Lambda$ hypernuclei below $\Xi$-instability and with 
$10<Z_{core}<120$, considering various choices for the unknown interaction channels such as VY0, VY1 and VY2 (Table~\ref{table:VSX}).
As expected, the largest corrections come from the unknown $\Lambda\Xi$ channel (VY1), and the $\Xi\Xi$ channel (VY2) has almost no impact on the
number of hypernuclei below S$_\mathrm{inst.}$.
This latest result is rather expected since the $\Xi\Xi$ interaction can occur only if the number of $\Xi$ at the onset threshold is at least 2, which rarely occurs.

\section{Conclusions}
\label{sec:conclusions}

In this work we have presented the first extensive microscopic exploration of the nuclear chart along the strangeness dimension where the competition between the $\Lambda$ and the $\Xi$ hyperons is consistently treated. 
The exploration of the nuclear chart as function of the strangeness number $S$ is performed by adding hyperons to a core ($A_{core},Z_{core}$) imposing either 
conserved total charge $Q$ or conserved proton number $Z$.
This study, which is a continuation of our previous work detailed in Ref.~\cite{Khan2015}, is performed using realistic and microscopically rooted non-relativistic energy functionals. 
In particular, we use in the $N\Lambda$ channel different functionals extracted from Brueckner-Hartree-Fock calculations with Nijmegen interactions. 
These effective interactions, fitted on all the available phase shifts, cover our present uncertainty on the interaction at least at low density, and have been successfully confronted to hypernuclear data in the past.
We have proposed a phenomenological extension of these potentials to the whole baryonic octet.
The experimental data on single and double $\Lambda$ hypernuclei are used to constrain the $N\Lambda$ interaction, and the mean-field analysis of the "Kiso" event is performed along the line 
proposed in Ref.~\cite{Sun2016} to determine the $N\Xi$ interaction.

Starting from a non-strange $(A,Z)$ core and adding $\Lambda$ hyperons, we have shown that the quasi-totality of the hypernuclei present an instability towards the decay into $\Xi$ hyperons before the strangeness drip-line is met. 
The strangeness instability threshold increases by step with the mass of the system due to shell effects.
It is approximately constant at a given $Q$ for stable $(A_{core},Z_{core})$ cores. 
A clear Coulomb effect is present, with $\Xi^0$ appearing in the proton rich side of the nuclear chart, and $\Xi^-$ for the majority of hypernuclei (at conserved charge $Q$). 
At conserved charge $Q$, the onset of the first $\Xi^{0,-}$ corresponds to the crossing between the $\Lambda$ and the neutron or proton chemical potentials.
We also show the impact of the different interacting channels on the results. 
The numerical value of the instability threshold largely depends on the $N\Lambda$ and $N\Xi$ interaction model, which are the most important channels.
The $\Lambda\Xi$ interaction has however a non-negligible impact: it can modify the number of pure-$\Lambda$ hypernuclei by 30-40\%.
Finally, the $\Xi\Xi$ interaction channel has almost no impact on the position of the $\Xi^{0,-}$ instability, and therefore on the number of pure-$\Lambda$ hypernuclei.
It seems rather weakly impact multi-strange hypernuclei.
In all cases the opening of the $\Xi$ channel reduces the number of bound pure $\Lambda$-hypernuclei that we have previously estimated~\cite{Khan2015} by a factor of approximatively 1/3-1/2, to
be about 200 000-300 000 hypernuclei.

The detailed characteristics of multi-hypernuclei along the nuclear chart, as well as their excited states, will be addressed in a future study.
We also plan to include $\Sigma$ and $\Omega$ hyperons and perform a sensitivity study on their largely unknown coupling, in order to further asset the possible model dependence of the results.

\begin{acknowledgments}
This work was partially supported by the SN2NS project ANR-10-BLAN-0503, by New-CompStar COST action MP1304, and by the IN2P3 Master Project MAC.
\end{acknowledgments}

\end{document}